\documentclass{IEEEtran}
\IEEEoverridecommandlockouts
\usepackage[normalem]{ulem}
\usepackage{amsmath,amsfonts}
\usepackage{xcolor}
\usepackage{mathrsfs}

\usepackage{array}
\usepackage{textcomp}
\usepackage{longtable}
\usepackage{nameref}

\usepackage{subcaption}
\usepackage{stfloats}
\usepackage{url}
\usepackage{verbatim}
\usepackage{graphicx}

\usepackage[ruled]{algorithm2e}
\usepackage{hyperref}
\usepackage{url}
\usepackage{cite}

\usepackage{multirow}

\usepackage{subcaption}

\usepackage[font=small,skip=0pt]{caption}
\usepackage[justification=centering]{caption}

\newcommand{\name}{DESTinE\xspace}

\usepackage{array}

\usepackage{tabularray}

\begin{document}

\title{Scalable Discrete Event Simulation Tool for Large-Scale Cyber-Physical Energy Systems: Advancing System Efficiency and Scalability

\thanks{This work was supported by the US Department of Energy 
under award DE-CR0000018 and the National Science Foundation under grant 2220347.}

}
\author{Khandaker Akramul~Haque\IEEEauthorrefmark{1}, \IEEEmembership{Graduate Student Member, IEEE},
Shining~Sun\IEEEauthorrefmark{1}, \IEEEmembership{Student Member, IEEE},
Xiang~Huo\IEEEauthorrefmark{1}, \IEEEmembership{Member, IEEE},
Ana~E.~Goulart\IEEEauthorrefmark{2},
\IEEEmembership{Member, IEEE}, and Katherine~R.~Davis\IEEEauthorrefmark{1},
\IEEEmembership{Senior Member, IEEE}\\
\IEEEauthorblockA{%
    \IEEEauthorrefmark{1}Department of Electrical and Computer Engineering, Texas A\&M University, College Station, TX, USA \\
    \IEEEauthorrefmark{2}Department of Engineering Technology and Industrial Distribution, Texas A\&M University, College Station, TX, USA
  }
}
\maketitle
\thispagestyle{plain}
\pagestyle{plain}

\begin{abstract}
Modern power systems face growing risks from cyber-physical attacks, necessitating enhanced resilience due to their societal function as critical 
infrastructures. The challenge is that defense of large-scale systems-of-systems requires scalability in their threat and risk assessment environment for cyber-physical analysis including cyber-informed transmission
planning, decision-making, and intrusion response. Hence, we present a scalable discrete event simulation tool for analysis of energy systems, called \name. The tool is tailored for large-scale cyber-physical systems, with a focus on power systems. It supports faster-than-real-time traffic generation and models packet flow and congestion under both normal 
and adversarial conditions.  Using three well-established power system synthetic cases with 500, 2000, and 10,000 buses, we overlay a constructed cyber network employing star and radial topologies. Experiments are conducted to identify critical nodes within a communication network in response to a disturbance. The findings are incorporated into a constrained optimization problem to assess the impact of the disturbance on a specific node and its cascading effects on the overall network. Based on the solution of the optimization problem, a new hybrid network topology is also derived, combining the strengths of star and radial structures to improve network resilience. Furthermore, DESTinE is integrated with a virtual server and a hardware-in-the-loop (HIL) system using Raspberry Pi 5. The performance of star, radial, and hybrid topologies is quantified under standalone operation, virtual server integration, and HIL setup to evaluate scalability and network performance. Results are compared for accuracy with the Common Open Research Emulator (CORE). The results show that \name is efficient and scalable for large-scale  test cases. These findings highlight \name's potential for real-time applications in large-scale cyber-physical systems.
\end{abstract}


\section{Introduction}

\IEEEPARstart{I}{n} recent years,
the resilience of power systems has been increasingly challenged by both natural disasters and deliberate incidents~\cite{Resiliency_davis2023}. While the risks of natural events such as hurricanes, earthquakes, and floods to power critical infrastructures have existed for a long time, the risks of intentional cyber-attacks are newer and growing. Concerns about intentional threat actors have raised awareness and interest for stakeholders to understand
the vulnerabilities and 
mitigate the risks in their systems.  Such threats have the potential to disrupt power supply, damage infrastructure, and cause extensive operational and financial setbacks. A notable example 
is the series of coordinated cyber-attacks on the Ukrainian power grid in 2015 and 2016 \cite{Assante2016}. These incidents not only underscore the susceptibility of critical infrastructures to cyber threats but also demonstrate their vulnerabilities during severe disruptions in service and security \cite{kshetri2017hacking}. 

The modern electrical grid is increasingly dependent on networks of cyber systems, such as in real-time data acquisition, control signal transmission, and 
decision-making assistance for both human operators and automated processes. 
While this integration 
has enhanced operational efficiency and responsiveness, it also introduces new challenges. The interconnected nature and complexity of cyber infrastructure create numerous potential vulnerabilities, exposing the grid to a variety of cyber threats. Malicious attacks targeting 
cyber components can disrupt data transmission, compromise control systems, and undermine the integrity of grid operations~\cite{Shining2024gridresponder}. As a result, it is critical to develop a comprehensive understanding of the interdependencies between cyber and electrical infrastructure to effectively mitigate these risks and safeguard the grid from evolving cyber threats \cite{dominguez2012reliability, kwasinski2020modeling}.

Cross-validation and comparison of proposed modeling and analysis approaches for cyber-physical systems (CPSs) are challenging due to the lack of standardized, reproducible models. Real-world models, particularly for grid cyber systems, are difficult to obtain as they are
sensitive
and rarely shared. A practical alternative is to utilize realistic reference models. For electrical power systems, several reference models are available, such as the 
widely used synthetic electric grids~\cite{tamu_repo}, but these are only for the phyiscal power system side, and they lack crucial cyber models, 
cyber-physical substations, and control architectures.
Realistic cyber-physical test cases are scarce, with most models limited to distribution systems and/or small 
test cases. 
Early efforts in cyber-physical transmission system modeling, such as the 8-substation CyPSA model \cite{weaver2016cyber} 
laid the groundwork for larger realistic cyber-physical grid models to be developed, like the
synthetic cyber-physical 2000-bus system 
\cite{wlazlo2019cyber}.  
Recently, more techniques have emerged for generating synthetic networks. A recent approach in \cite{liu2023gnn} utilizes graph variational autoencoders and recurrent neural networks to generate synthetic networks while protecting sensitive grid topology. However, this method has been tested only on small-scale test cases. Consequently, its suitability for generating large-scale cyber-physical models remains uncertain.

\subsection{Literature Review and 
Purpose of 
\name
}
In this work, we build upon the recent scalable cyber-physical model generation tool established in~\cite{samantha2024} and previous research on the comparison of graph theory metrics with simulations~\cite{haque2024graphtheoryvstimedomain} to enhance the scalability of cyber environment simulation with a discrete-event simulation approach in a solution called \name. By integrating graph-theoretic metrics, referred to as the network analysis matrix in this study, with high-fidelity discrete-event simulation tool SimPy~\cite{SimPy}, \name offers a comprehensive framework for scalable discrete event simulation. 

Previous studies exist that demonstrate
the effectiveness of SimPy in small-scale network simulations, where SimPy is
compared to established tools such as the Objective Modular Network Testbed in C++ (OMNeT++), Network Simulator 3 (NS-3), and the Java In Simulation Time / Scalable Wireless Ad hoc Network Simulator (JiST/SWANS) \cite{weingartner2009performance}. Moreover, a network slicing simulator developed with SimPy has been utilized to evaluate the performance of base station setups within 5G networks \cite{syed2022performance}. Additionally, SimPy has been employed to facilitate the integration of cloud and fog computing frameworks for future 5G networks, addressing the growing demands of mobile devices \cite{tinini20205gpy}. In this study, we 
analyze a small network using the Common Open Research Emulator (CORE) and compare the results with those obtained from our 
simulation tool in Section \ref{result} \cite{ahrenholz2008core}.


%
Existing cyber-physical system simulation tools face trade-offs in accuracy, scalability, and adaptability. CyPhySim, based on Ptolemy II, supports various simulation techniques but faces 
computational complexity challenges%
~\cite{lee2015modeling}. CPS-Sim integrates Matlab/Simulink and QualNet/OMNeT++ but encounters synchronization bottlenecks \cite{suzuki2018cps}. COSSIM, an open-source system-of-systems simulator, includes power estimation and security analysis but is computationally expensive \cite{brokalakis2018cossim}. RTCPS, a real-time testbed for power system cybersecurity, automates cyberattack testing but is limited to power grid applications~\cite{nguyen2024real}. In~\cite{sahu2021design}, authors present the design and
evaluation of a cyber-physical power system
testbed called Resilient
Energy Systems Lab (RESLab) and offer detailed
comparisons and analysis of different cyber-physical power system testbeds including their simulation, emulation, and hardware-in-the-loop features.
Network emulation tools, such as Mininet, offer a flexible platform for modeling and testing cyber-physical interactions in distribution networks \cite{10080763}. However, Mininet often encounters challenges when integrating hardware-in-the-loop (HIL) systems \cite{flauzac2019mininet}. These limitations underscore the need for a more scalable and efficient CPS simulation framework.


The proposed method in \name enables a topology-informed security assessment, allowing the identification of critical nodes and the evaluation of their performance and reliability under cyber threats in CPSs, particularly  in power grids. Based on the assessment score, the number of affected substations, as well as the specific utility impacted during the adversarial event, we categorize the severity of impact to the overall network function due to cyber-node loss into six distinct levels. The classification categories are outlined in Section~\ref{result}, which subsequently guide potential network reconfiguration.

\name's
integrated approach enables a comprehensive evaluation of network vulnerabilities and facilitates the optimization of defensive strategies within cyber-phyiscal smart grid environments, ensuring a more resilient and secure infrastructure. Moreover, the proposed approach mitigates the computational complexity associated with large-scale networks. We have implemented a Denial of Service (DoS) attack at various locations within the CPS (Section \ref{result}) using the simulation tool. 
DoS attacks are chosen to study because they directly impact network connectivity and data flow, making them highly relevant to cyber-physical energy systems. They are also one of the most common threats, and their effects on packet loss and network degradation can be systematically quantified, making them a suitable initial case study for DESTinE \cite{eliyan2021and}. Additionally, a DoS attack in our framework may be interpreted as the physical removal of a node, which can also represent other cyber threats that disrupt network stability. Future work will focus on incorporating additional adversarial events to develop a threat-agnostic analysis framework.

\subsection{Comparing Simulation and Emulation}

Simulation and emulation are essential techniques in network analysis. Each has a unique role in system modeling and testing, especially for large-scale cyber-physical power system applications. Simulation relies on mathematical models and algorithms to represent component interactions over time, therefore providing a simplified view of the system without recreating the exact environment. This makes simulation suitable for predictive analysis, evaluating performance under hypothetical scenarios like extreme loads, equipment failures, or cybersecurity attacks. For large-scale power systems, simulation enables researchers to study vulnerabilities, resilience, and scalability.

Emulation, on the other hand, replicates one system’s functionality within another system, creating a realistic environment by running actual softwares and protocols. This approach is particularly useful for testing model accuracy and compatibility, verifying operational reliability, and evaluating network behavior in near-real-world conditions. Emulation allows software and components to interact as if in their native environment, enabling a realistic and detailed assessment of system responses, such as intrusion detection and recovery. While emulation generally offers higher fidelity by closely replicating the operational environment, simulation scales more efficiently by reducing physical constraints, making it ideal for theoretical studies and performance predictions. Both techniques offer valuable insights into system behavior, supporting cybersecurity objectives in large-scale systems like power grids by identifying vulnerabilities, evaluating performance, and testing resilience under varied conditions.

Importantly, a simulator can be adapted for use as an emulator by integrating real network traffic and device interactions into the simulation framework, thus enabling a more realistic environment while maintaining the inherent flexibility of a simulation. This hybrid approach is achieved by modifying the simulator to accommodate live data inputs or to execute actual software components within the simulated environment. In the context of cyber-physical power systems analysis, simulators can be enhanced to process real protocol exchanges or to interface directly with networked devices, allowing system behaviors and responses to be observed with greater accuracy. Through the incorporation of real-world data and/or live network configurations, a simulated yet realistic test environment is created, which can closely mirror the actual system. This facilitates compatibility checks, intrusion detection assessments, and operational stress testing. Such an approach retains the resource efficiency and adaptability characteristics of simulation while reaching the high fidelity associated with emulation, offering a valuable solution when both scalability and realism are required.



\subsection{DESTinE: Overview of Hybrid Test Environment and Key Contributions}
In this study, a hybrid test environment and Scalable \underline{D}iscrete \underline{E}vent \underline{S}imulation \underline{T}ool \underline{in} \underline{E}nergy (\name) are designed to effectively combine emulation capabilities with simulation functionality. Hybrid tools like \name can replicate high-level and underlying processes, timing, and interactions of real-world systems, allowing users to interact with a virtual environment as if it were fully operational. Through process-based modeling, \name would 
be able 
to 
emulate network communication, manage real-time delays, and replicate the behavior of hardware components such as microcontrollers and sensors, creating a dynamic testbed that behaves similarly to an actual deployment.

In addition to the aforementioned  capabilities, DESTinE can generate network traffic patterns that simulate various operational scenarios, including normal communication loads and potential cyber-adversarial conditions. The network packets generated by DESTinE can vary in size and timing to closely mirror the real-world complex network traffics. This traffic generation can be customized to mimic specific protocols, device behaviors, or network conditions, enabling accurate analysis of system performance, security, and resilience. By providing a high-fidelity hybrid test environment that incorporates real-world software  interactions and emulated hardware responses, DESTinE serves as a powerful tool for assessing the functionality and robustness of CPSs.




In summary, 
\name addresses key scalability challenges in analyzing the cyber side of large-scale power systems.
While the physical side can be handled using tools like PowerWorld (PW) \cite{klump2002powerworld}, the proposed simulation tool focuses on enhancing the cyber aspect of the power system, offering a robust network functionality assessment approach 
for
large-scale power system analysis. In 
contrast, the developed approach
offers scalability in the analysis of the cyber domain within cyber-physical power systems. 
The contributions of this paper 
are as follows:
\begin{itemize}

    \item  We introduce a scalable discrete event simulation tool for use in the analysis of energy systems, called \name that effectively mimics aspects of emulation. \name replicates both high-level and underlying processes, timing, and interactions of real-world systems, allowing users to interact with a virtual environment as if it were fully operational.  Through process-based modeling, \name can emulate network communication, manage real-time delays, and replicate the behavior of hardware components such as microcontrollers and sensors. 
    
    \item \name is evaluated on three large-scale synthetic power systems with 500, 2000, and 10,000 buses, where the proposed approach is applied to simulate their cyber networks based on star and radial topologies. 
    
    \item A comparative analysis of various network metrics is conducted to identify critical nodes. We evaluate these metrics against nodes identified as critical based on average delay measurements obtained from the simulation tool when each system is under DoS attack.  Following the DoS attacks on all utilities, an optimized list of critical utilities is generated, as detailed in Section~\ref{result}. 
    \item \name's optimization model further classifies the utilities into six risk categories based on the impact of the DoS attack. 
    A constrained optimization model is developed to generalize this approach,
    detailed in Sections 
    \ref{network analysis} and \ref{result}.
    \item The approach developed in \name improves the efficiency of identifying network bottlenecks and vulnerabilities by modeling packet flow and congestion under various conditions, including peak traffic and cyber adversarial events.
    %
    \item After the optimizer has classified the utility nodes based on severity, a hybrid model is proposed and implemented  by restructuring the substations connected to the utility in either a radial or star configuration. Additionally, a comparative analysis of the hybrid model is conducted to evaluate its effectiveness.
    \item The simulator is further integrated with a virtual server and subsequently with a hardware-in-the-loop (HIL) setup using a Raspberry Pi 5 to emulate a real-time network environment \cite{rpi5}. Furthermore, a comparative analysis is performed to evaluate the efficacy and scalability of the proposed hybrid simulation framework. 
    
\end{itemize}

The potential applications of \name are substantial. By synchronizing with real-world time or adjusting time scales, the simulation accurately mimics energy transactions, grid responses, and communication delays, enabling realistic and comprehensive testing of system performance. Furthermore, DESTinE is capable of modeling interactions between distributed energy resources (DERs), such as solar, wind, and storage systems \cite{huo2024review}, and centralized energy management systems (EMSs), optimizing energy dispatch and load balancing for next-generation EMS \cite{sahu2023design}.
Hence, \name can facilitate the study of 
multi-time-scale coordination in transmission and distribution systems by simulating communication between DERs and the EMS, ensuring efficient energy flows and improving grid performance
under various conditions. Additionally, the tool 
should help identify control challenges associated with high penetration of renewables.

The rest of the paper is structured as follows. Section \ref{framework} outlines the framework used to integrate large-scale power system models with the discrete event simulator. Section \ref{network analysis} details the network analysis matrix  used to identify key nodes. Section \ref{result} presents the results under both normal operations and during a DoS attack. It also discusses the generation of hybrid a topology and evaluates the performance of \name using both a virtual server and a HIL setup, thereby enhancing its scalability and real-world applicability. Finally Section V concludes the paper and highlights the future potential and applications of \name.

\section{Framework of DESTinE} 
\label{framework}
\begin{figure}[!t]
\centerline{\includegraphics[scale=0.72]{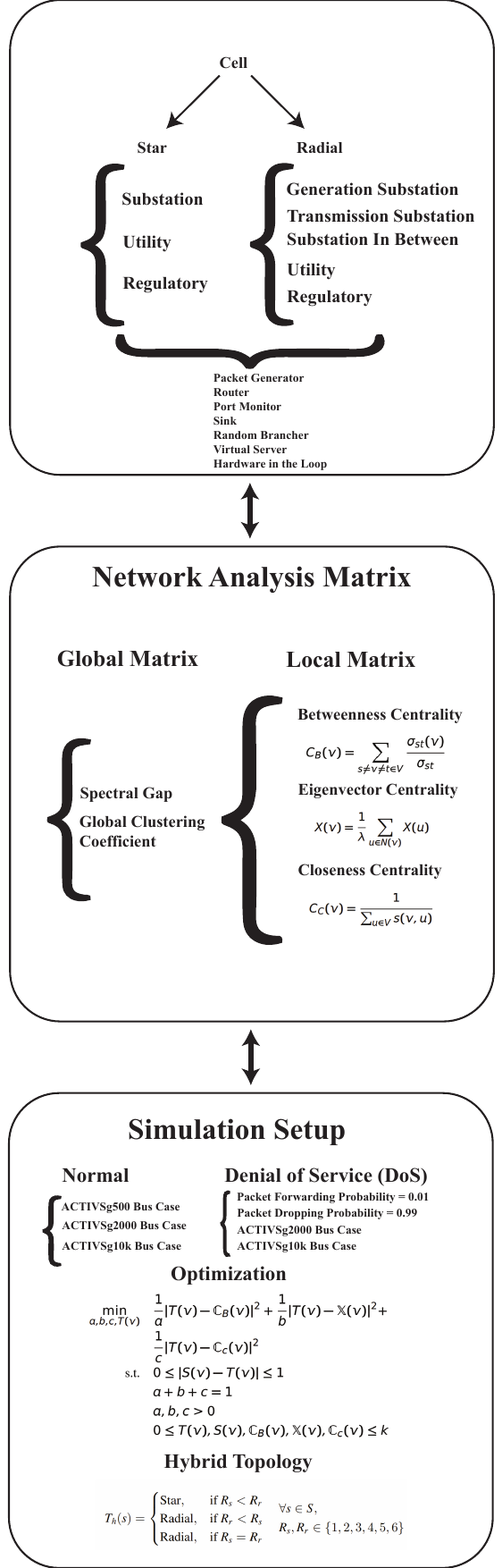}}
\captionsetup{justification=justified}
\vspace{3mm}
\caption{Diagram illustrating the framework of the simulation tool, with each square representing a distinct module, highlighting their interconnected roles in the overall simulation process. }
\label{fig:framework}
\end{figure}


The framework of \name consists of three main modules: (1) The first module is a discrete event simulator, which facilitates the simulation of various network scenarios. (2) The second module incorporates global and local matrices, where the global matrices provide a high-level overview of a large network, enabling the generation of subgraphs, while the local matrices indicate the importance of nodes/buses based on their proximity to other critical nodes. Collectively, we refer to these as the network analysis matrices. (3) The third module addresses the simulation setup under normal operating conditions and outlines the adjustments made during a DoS disturbance. The framework is given in Fig. \ref{fig:framework}. 
In this section, we will provide a detailed elaboration of the first module, while the second and third modules will be briefly introduced, with a more in-depth analysis to follow in subsequent sections. 

The simulation tool accepts input in the form of a JavaScript Object Notation (JSON) object that represents all substations, utilities, and regulatory units, with each element modeled as a router. The models and the model generation
algorithm are detailed in~\cite{samantha2024}. Some routers are connected to a packet generator, while others connect to a sink, based on the connections specified in the JSON object. Looking ahead, we plan to introduce greater flexibility in the tool to accommodate various types of inputs, enhancing its adaptability for future simulations. 


\begin{figure}[!t]
    \centering
    \begin{subfigure}{0.4\textwidth}
         \centering
         \includegraphics[width=\textwidth]{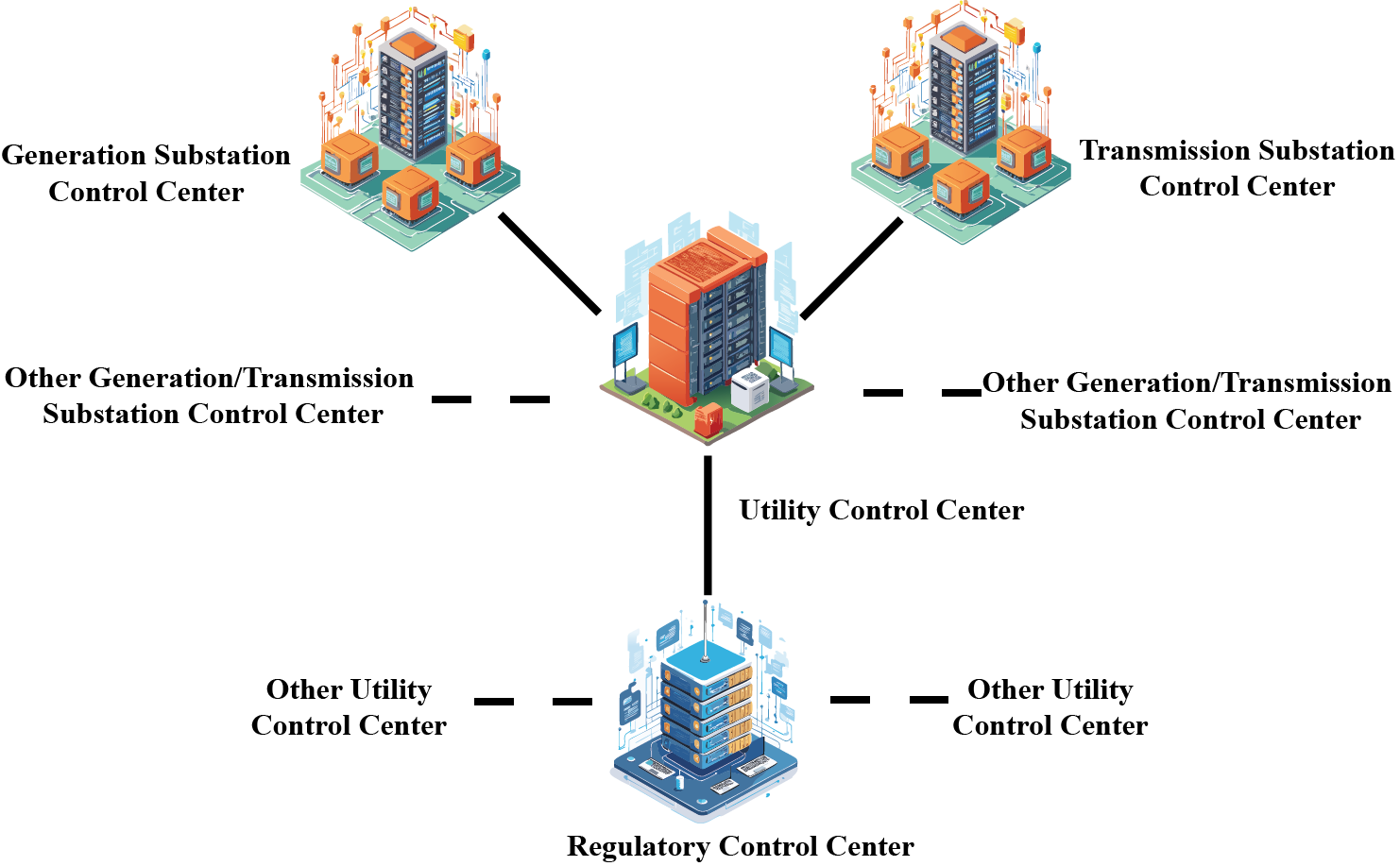}
         \caption{}
     \end{subfigure}\hfill
     
     \begin{subfigure}{0.35\textwidth}
         \centering
         \includegraphics[width=\textwidth]{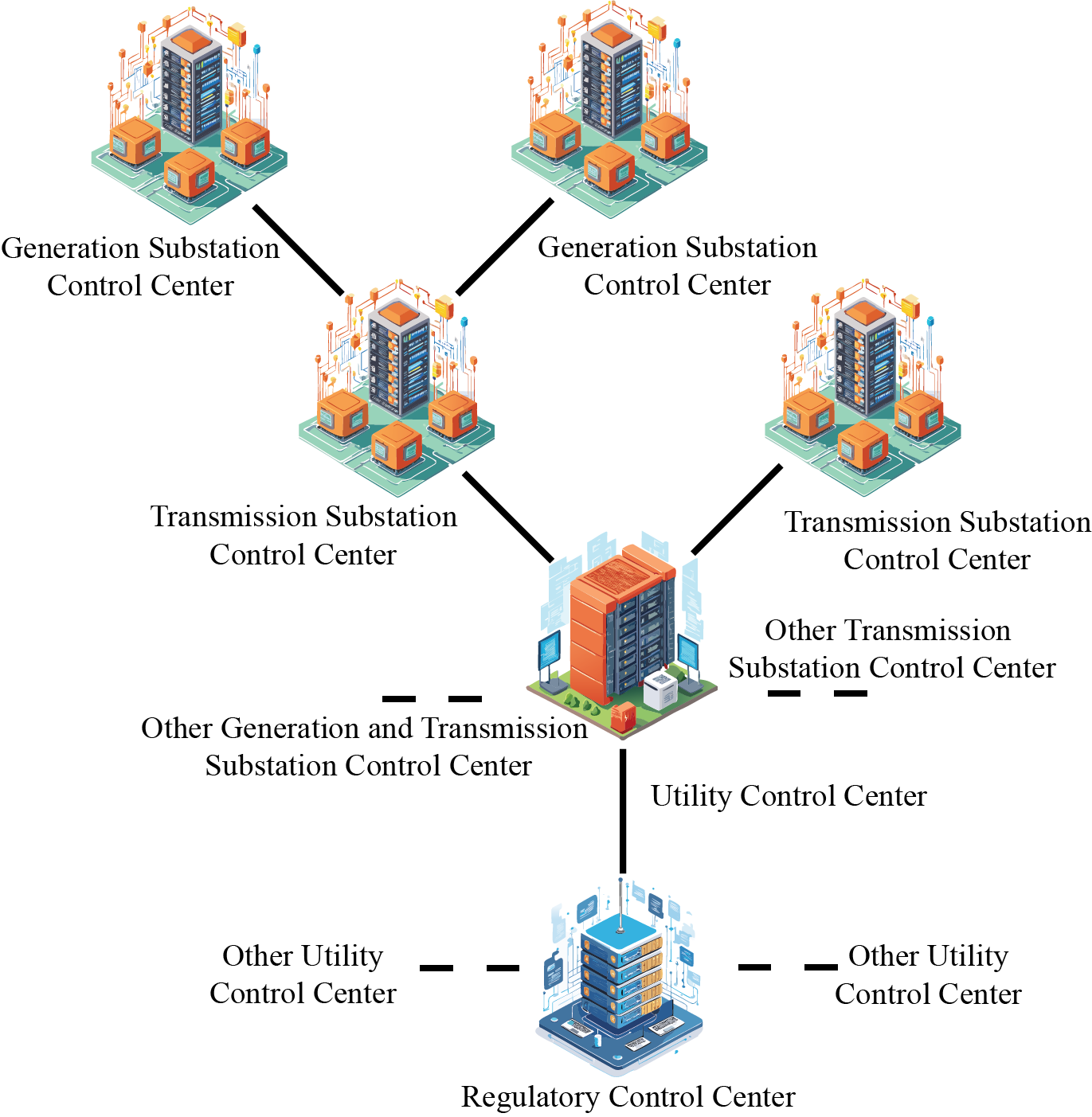}
         \caption{}
     \end{subfigure}
     \captionsetup{justification=justified}
    \caption{Structure of cell (a) Star topology and (b) Radial topology }
    \label{fig:cell_structure}
\end{figure}

\begin{figure}[!t]
\centerline{\includegraphics[scale=0.7]{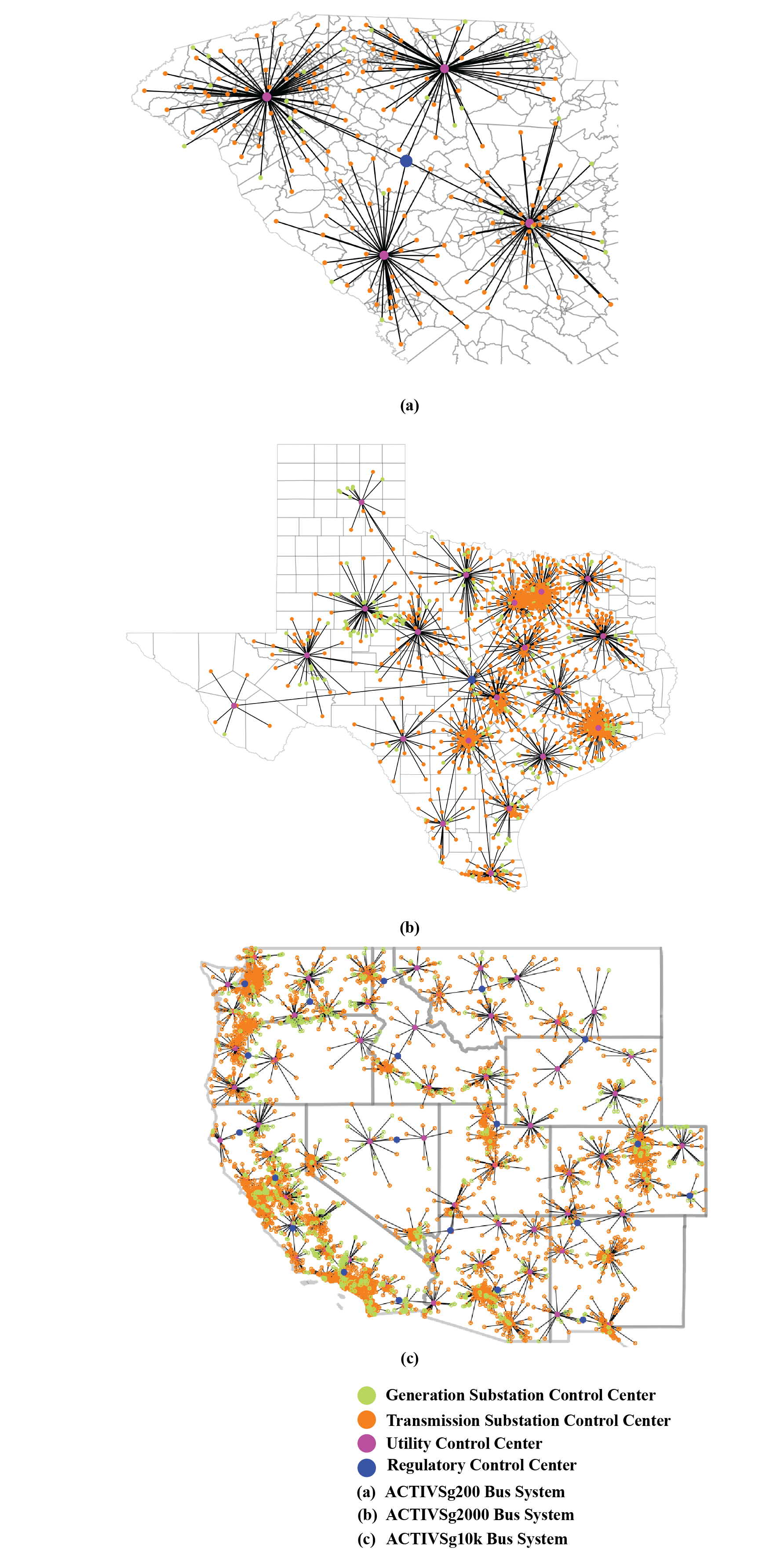}}
\captionsetup{justification=justified}
\caption{The connection of network elements visualized in \name in a star topology in relation to the ACTIVSg500-bus system, the ACTIVSg2000-bus system, and the ACTIVSg10k-bus system \protect\cite{samantha2024}. }
\label{fig:star}
\end{figure}

\begin{figure}[!t]
\centerline{\includegraphics[scale=0.7]{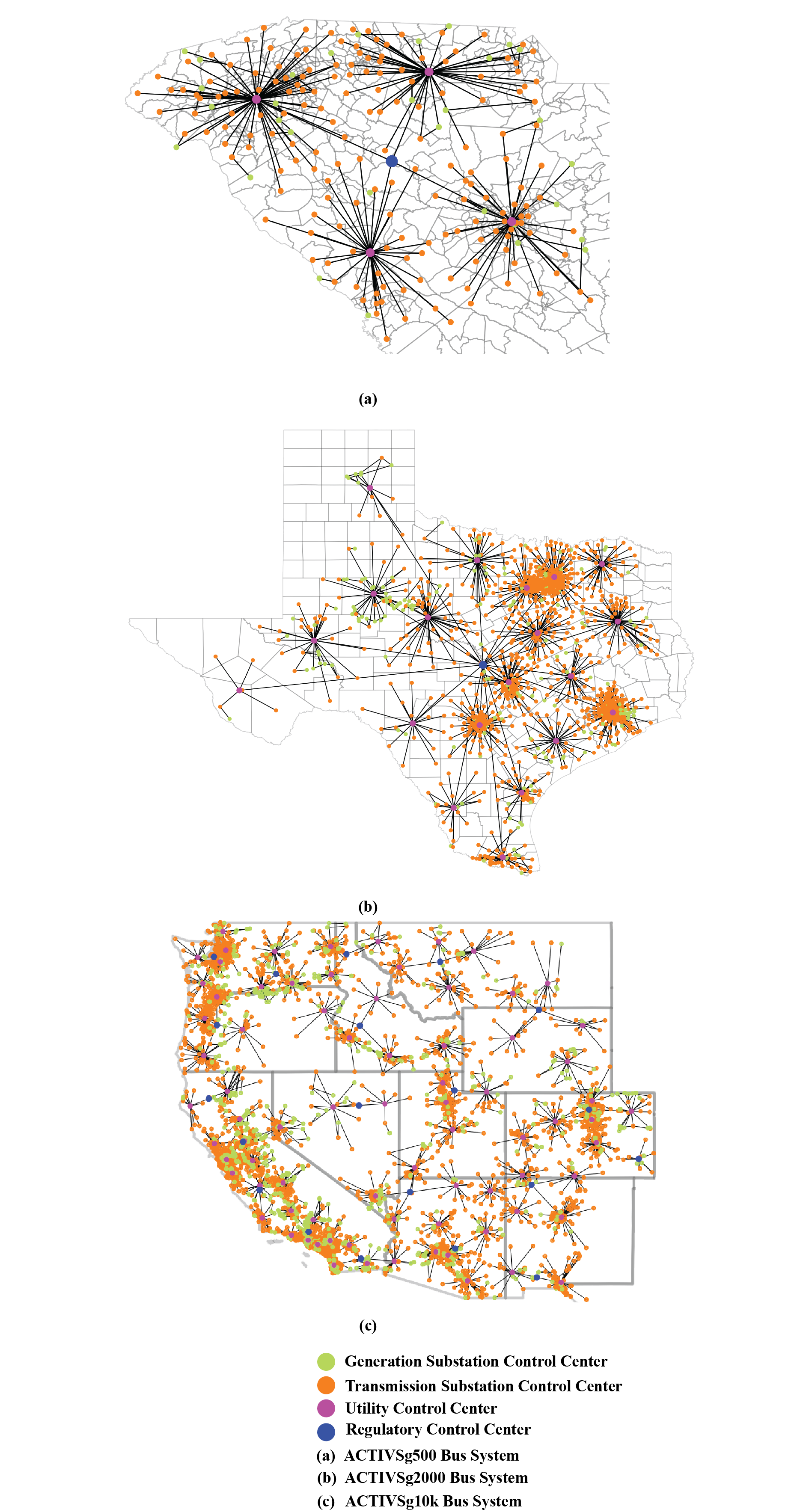}}
\captionsetup{justification=justified}
\caption{The connection of network elements visualized in \name in a radial topology in relation to the ACTIVSg500-bus system, the ACTIVSg2000-bus system, and the ACTIVSg10k-bus system \protect\cite{samantha2024}.}
\label{fig:radial}
\end{figure}

\subsection{Discrete Event Simulator}

The developed discrete event simulator tool \name uses SimPy, a Python-based process-driven discrete-event simulation framework, renowned for its scalability and efficiency. Python's generator functions underpin this framework by allowing the efficient iteration of data, yielding values one at a time with the \emph{yield} keyword, rather than returning all results simultaneously. This enables on-demand computation, reducing memory usage by maintaining only one value in memory at a time. Generators retain their state between yields, resuming execution when the next value is requested, a feature that SimPy leverages to simulate processes in discrete-event simulations. In this context, processes yield control during events like timeouts or resource waits, enabling intuitive modeling of asynchronous behavior. SimPy’s architecture, built on Python’s generator functions, facilitates flexible and efficient modeling of concurrent processes through coroutines, significantly reducing memory and CPU overhead. This lightweight design allows SimPy to handle numerous concurrent events without performance degradation. Its event-driven model processes only active events, therefore avoiding unnecessary computations, making it ideal for large-scale system simulations. Moreover, SimPy’s efficient event queue ensures proper sequencing of events, and scaling simulations while maintaining performance by minimizing computational load \cite{SimPy}.

\subsection{Case Study Descriptions}

In the context of power systems, characterized by their non-linear behavior and computational complexity, the deployment of SimPy is promising in modeling the cyber-physical environment by simulating the cyber network. To this end, this study models three large-scale power system cases, each overlaid with a cyber network: the ACTIVSg500-bus system \cite{birchfield2017SouthCarolina500}, the ACTIVSg2000-bus system \cite{birchfield2016Texas2k}, and the ACTIVSg10k-bus system \cite{birchfield2018WestUS10k}. Each power system is analyzed using two distinct network topologies, i.e., star and radial, providing valuable insights into the interactions between the cyber and physical layers across different network structures. This approach aims to enhance the understanding of cyber-physical dynamics in large-scale power systems, contributing to more scalable and efficient system designs. While this study focuses on star and radial topologies, 
the framework can be adapted to simulate various network topologies and configurations, extending its applicability beyond star and radial topologies. 
This generalizability makes it a versatile tool for analyzing cyber-physical interactions across diverse network structures, aiding in the design and evaluation of large-scale power systems with varying operational and security requirements.  


Each power system in this study comprises several types of substations with distinct functions. These include generation substations, which house generating units such as generators connected to turbines; transmission substations, which facilitate power transmission without generation capabilities; utility units, responsible for the distribution of power to consumers; and regulatory units, also known as load dispatch centers, which oversee the control and management of power transmission. The specific number of generation substations, transmission substations, utility units, and regulatory units for each power system case is detailed in Table \ref{tab1}.

The star and radial topologies represent distinct network configurations chosen for this study due to their simplicity and scalability, which allow for meaningful comparisons across different system sizes. In a star topology, all nodes are directly connected to a central node, such as a utility control center, making this central node crucial for communication. A failure at the central node results in complete network disconnection. On the other hand, the radial topology follows a hierarchical structure, with the central node connecting to secondary nodes, which, in turn, connect to additional nodes. As discussed in \cite{huang2020planet}, this configuration transforms the network graph into a spanning tree, making it particularly robust and scalable for transmission networks \cite{short2003electric}. The radial topology enables exploration from a root node, facilitating the visualization of hierarchical relationships and identification of key nodes within the network. Moreover, framework related to star topology has been developed due to their prevalence and significance in power system architectures \cite{star_frame}.

While real-world systems may deviate from these idealized structures, the star and radial topologies are selected to provide a homogeneous basis for the comparison of large-scale cyber-physical systems with different system sizes. Their relative simplicity makes them easy to implement for scalability testing, offering insights into the performance of large-scale systems without compromising the generality of the analysis. Therefore, these topologies are used to demonstrate key aspects of network resilience and communication efficiency. Moreover, the analysis remains applicable to more complex or hybrid configurations, ensuring that the findings are not limited to these specific types of topologies.

\begin{table}[!t]
\centering
\caption{Number of entities of different cases.}
\label{tab1}
\begin{tblr}{
  width = \linewidth,
  colspec = {Q[219]Q[237]Q[246]Q[237]},
  hlines,
  vlines,
}
\textbf{Entity } & \textbf{ACTIVSg500 bus system} & \textbf{ACTIVSg2000 bus system} & \textbf{ACTIVSg10k bus system}\\
Generation Substation & 31 & 188 & 851\\
Transmission Substation & 177 & 1062 & 3921\\
Utility Unit & 4 & 20 & 80\\
Regulatory Unit & 1 & 1 & 20
\end{tblr}
\end{table}

\subsubsection{Procedure for setting up star and radial topology in the discrete event simulator}

To establish a star topology in a discrete event simulator, the fundamental unit, referred to as a ``cell'', must first be defined. This cell includes a generation substation control center, a transmission substation control center, a utility control center, and a regulatory control center (Fig.  \ref{fig:cell_structure}). The creation of this cell is driven by five key SimPy environment components, implemented using object-oriented programming. These components, 
the \textbf{Packet Generator}, \textbf{Router}, \textbf{Port Monitor}, \textbf{Sink}, and \textbf{Random Brancher}, are defined as classes within the DESTinE framework. Additionally, two new classes, \textbf{Virtual Server} and \textbf{Hardware-in-the-Loop (HIL)}, have been introduced to enable integration with external applications. Unlike the previously mentioned five classes derived from SimPy, these two classes are independently designed but are generalizable and can interact with SimPy’s environment when required. Their addition enhances DESTinE’s flexibility, allowing seamless interaction with virtual servers and physical hardware, thereby extending its applicability beyond standalone simulations.



\begin{itemize}
        \item \textbf{Packet Generator}: The Packet Generator is responsible for creating network packets at specified time intervals based on a user-defined inter-arrival distribution. Each packet carries information such as arrival time, size, and source-destination IDs. Once generated, packets are sent to the next network element, router.
        \item \textbf{Router}: 
         The Router in DESTinE manages packet forwarding and queuing behavior. It receives incoming packets, checks queue capacity (in bytes), and drops packets if the limit is exceeded, simulating congestion. The router forwards packets at a specified bitrate, incurring transmission delay proportional to packet size, and passes the packet to the next connected element once transmission completes.
        \item \textbf{Port Monitor}: The Port Monitor observes the router's queue status at regular intervals, recording metrics such as queue length or byte size to capture the system's state for performance evaluation. This data helps analyze bottlenecks, congestion levels, and overall network behavior during the simulation, providing insights into the router's efficiency and potential areas for optimization.
        \item \textbf{Sink}: The Packet Sink serves as the final destination where packets are collected and stored, recording packet-level statistics such as delays, inter-arrival times, and packet loss. This data enables detailed post-simulation analysis of network performance metrics, providing insights into the efficiency and behavior of the simulated network.
        \item \textbf{Random Brancher}: The Random Brancher directs packets to different downstream network paths based on assigned probabilities, simulating random routing decisions in a network. This functionality allows for branching flows into multiple paths, increasing the complexity and realism of the simulation by mimicking real-world network behavior.
        \item \textbf{Virtual Server}: The DESTinE tool is designed to connect with a virtual server using a Transmission Control Protocol (TCP) socket \cite{donahoo2009tcp}. In this implementation, a remote JavaScript-based Node.js server is utilized to facilitate communication. Since actual network packets are transmitted, standard packet capture tools such as Wireshark can be employed for traffic analysis, providing enhanced monitoring and security assessment capabilities \cite{lamping2004wireshark}.
        \item \textbf{Hardware-in-the-Loop (HIL)}: For the hardware-in-the-loop (HIL) setup, a Raspberry Pi 5 is integrated due to its cost efficiency and versatility as a single-board computer, making it well-suited for emulating real-world devices. In this configuration, a single Raspberry Pi 5 is utilized to simulate multiple hardware-in-the-loop components by leveraging its virtual ports. The DESTinE tool establishes a TCP socket connection with the Raspberry Pi, facilitating seamless data exchange and real-time system emulation.
        
\end{itemize}

These classes are instantiated every time an object is created, with specific values initialized during object creation. Each class contains methods to perform essential functions, such as packet generation, packet forwarding, storing packets, and tracking packets received and lost. The creation of these objects follows a dynamic programming approach. Dynamic programming in this simulation framework enables efficient and adaptive instantiation of objects, tailored to fit the underlying network topology. Rather than statically defining connections or packet routes, dynamic programming allows for flexible object creation and management based on each network’s specific layout. For instance, in a star topology, packets flow dynamically from each substation control center (generation or transmission) to the utility control center, reflecting the hierarchical data aggregation typical in such configurations. This adaptable routing enables the utility control center to receive and process data from all connected substations efficiently. In radial topologies, packets are dynamically routed along linear or branching paths, ensuring data flows through each level until it reaches the utility control center. Thus, we aim to achieve  a flexible framework for modeling cyber-physical interactions across a variety of network structures, supporting comprehensive analysis of large-scale power systems.



\subsubsection{Star Topology}
In the star topology, all substation control centers, both generation and transmission, are interconnected with the utility control center, with the nearest utility control center selected based on geographic proximity. Within the simulator, each substation is associated with its respective packet generator, which connects to a router. These substation routers then link to utility routers, which are programmed to either store packets in the utility sink or forward them to the regulatory router with equal probability. The regulatory router subsequently stores the forwarded packets in the regulatory sink. An illustration of star topology is depicted in Fig. \ref{fig:star}.

\subsubsection{Radial Topology}

In a radial topology, the packet generator is linked to the generation substation, which is then connected in its communication infrastructure to the generation substation router. This router is further connected to the transmission substation router based on geographic proximity. Depending on proximity, the transmission substation router may also connect to other generation substation routers. The connection of the transmission substation router to the utility router and regulatory unit follows the same principles as those used in a star topology. Since the number of generation substations is less compared to transmission substations in all three cases, a large number of packet generators are also connected to transmission substations that lack generation substations. These packet generators are linked to the transmission substation router, which in turn connects to the utility router and eventually to the regulatory router. The connection of sinks to both the utility router and the regulatory router follows a similar pattern to that of the star topology.



For both the ACTIVSg500
and ACTIVSg2000-bus systems, there is only one regulatory unit, meaning all utility routers connect to a single regulatory router. In contrast, ACTIVSg10k-bus system has 20 regulatory units, so utility routers connect only to the regulatory router within the jurisdiction of the respective regulatory unit. This jurisdictional configuration aligns with the power system network connections detailed in \cite{birchfield2018WestUS10k}. For all of these cases, the port monitor variable will be used to monitor the routers at the utility control center and the regulatory control center, as they collect data from all substations. The radial topology is shown in Fig. \ref{fig:radial}.
Pseudo codes for generating star and radial topology are given in \nameref{appendix}.

\subsection{Network Analysis Matrix}
To analyze the cyber network of the power system test cases, we integrate both global and local matrices to gain a comprehensive understanding of the network's structure. The global matrix offers an overarching view, assessing the network's sparsity and its potential to be partitioned into smaller sub-networks. In contrast, the local matrix focuses on individual nodes or smaller network segments, evaluating characteristics such as node degree and shortest paths to determine the influence and significance of specific nodes within their immediate context. By combining these approaches, we achieve a thorough macro and micro-level analysis, providing a detailed perspective on the network's overall topology and the dynamics of its components. These analyses will be further elaborated in the following section.

\subsection{Dynamic Simulation of Cyber Network Routers: Methodology and Insights}

The event-driven model in our simulation efficiently manages network events, creating a streamlined environment tailored to simulate Layer 3 network behaviors according to the Open Systems Interconnection (OSI) model. In the SimPy framework, packets are generated based on an exponential distribution, with a mean packet size of 3.4 Megabytes \cite{bertsekas2021data}. The packet inter-arrival time also follows an exponential distribution, ensuring a realistic traffic flow in the network. As detailed in Table \ref{tab: simulation_params},  the simulation setup is designed with routers forwarding packets at an average rate of 2.2 packets per second.


\begin{table}[!t]
    \centering
    \scriptsize
    \caption{SimPy Simulation Environment: Setup and Parameters.}
    \label{tab: simulation_params}
    \begin{tabular}{lll} \hline
         Parameter & Probability Distribution & Value  \\ \hline
         Packet Size & Exponential & Mean = 3.4 Megabytes \\ 
         Packet inter-arrival time & Exponential & Max = 0.05 sec  \\ 
         Router's port rate  & Exponential & 2.2 packets/sec \\ 
         Sampling rate & Exponential & Max = 20 samples/sec \\ \hline
    \end{tabular}
\end{table}

\begin{table*}[!tbh]
\centering
\caption{Global Matrix Analysis of Power System Test Cases}
\label{tab: global matrix}
\begin{tblr}{
  width = \linewidth,
  colspec = {Q[94]Q[135]Q[142]Q[138]Q[146]Q[137]Q[142]},
  hlines,
  vlines,
}
\textbf{Global Values} & \textbf{ACTIVSg500 Bus system (Star Topology)} & \textbf{ACTIVSg500 Bus system (Radial Topology)} & \textbf{ACTIVSg2000 Bus system (Star Topology)} & \textbf{ACTIVSg2000 Bus System (Radial Topology)} & \textbf{ACTIVSg10k Bus System (Star Topology)} & \textbf{ACTIVSg10k Bus System (Radial Topology)}\\
Spectral Gap & 1.58E-02 & 1.57E-02 & 5.32E-03 & 5.32E-03 & 0 & 0\\
Global ClusteringCoefficient & 0 & 1.88E-02 & 0 & 1.08E-02 & 0 & 1.67E-02
\end{tblr}
\end{table*}

Each router, connected to utilities and regulatory units, is closely monitored during the simulation. Key metrics recorded include the router service time, which measures the delay from when a packet enters the router to when it exits. This service time is logged every 0.5 seconds, and the delay is calculated as the ratio of the queue waiting time to the number of packets processed. Queue waiting time represents the average duration each packet waits before being forwarded. Additionally, the simulation tracks the packet size and the total number of packets processed, providing insights into network traffic characteristics and performance.
Under normal conditions, this approach ensures that the router's queue capacity is sufficient to prevent packet loss, maintaining the integrity of the simulation. The detailed setup and analysis of a DoS attack scenario, which tests the resilience of this configuration, will be elaborated in Section \ref{result}.


\noindent \textbf{Remark 1:} 
Note that our simulation focuses exclusively on nodes functioning as routers within the cyber network, deliberately excluding overhead associated with these nodes to streamline computational requirements. Specifically, routing protocol overhead, packet processing delays, and control message exchanges typical in real-world routers are not considered. Additionally, we omit Layer 2 (data link layer) and Layer 1 (physical layer) communication overhead, such as Ethernet frame headers and physical layer signaling, which can influence packet size and processing times. The effects of individual devices, such as computers or IoT devices within private networks, which may introduce variability in traffic patterns and processing loads, are also excluded. The implications and impacts of these requirements are highlighted below:

\begin{itemize}
    \item Streamlined simulation and targeted insights: By excluding these detailed overheads, the simulation allows a focused examination of the primary interactions within the network. This approach provides clear and valuable insights into the system's broader behavior, particularly in understanding critical network interactions without being bogged down by granular details.
    \item Potential overestimation of performance: Omitting these factors may result in a slightly optimistic assessment of the network's performance and resilience, as real-world routers and communication links would experience additional processing delays and bandwidth consumption due to these overheads.
    \item Negligible impact on accuracy for targeted applications: While the exclusion of overhead may lead to minor discrepancies, these effects are not significant for our targeted analysis because processing delays from these omitted factors are typically small. Thus, the simulation remains highly effective for assessing critical scenarios, such as node failures or attack impacts on network resilience.
    \item Suitability for large-scale scenarios: The reduction in computational complexity makes the approach particularly well-suited for large-scale network analyses, where the inclusion of every detail could significantly slow down processing and limit the scope of the study.
\end{itemize}
This balance between simplification and realism ensures the simulation is both efficient and insightful, providing a practical framework for evaluating the key dynamics of cyber-physical systems.


\section{Network Analysis Matrix and Optimization} 
\label{network analysis}

The network analysis matrices include 1) the global matrix that offers a comprehensive view of network connectivity, highlighting patterns like sparsity and identifying potential partitioning opportunities, and 2) the local matrix that focuses on individual nodes and their specific connections. Detailed explanations are provided below.


\subsection{Global Matrix}


To gain an initial understanding of the cyber network structure for each power system, we employ a global matrix analysis, focusing on two key parameters: the Spectral Gap and the Global Clustering Coefficient.

\textbf{Spectral Gap}: This parameter is defined as the first nontrivial eigenvalue of the normalized Laplacian matrix. It serves as a numerical indicator of whether a graph can be divided into subgraphs. A small spectral gap suggests the possibility of division, with a value approaching 0 indicating that a perfect graph division is feasible \cite{chung1997spectral}. The equation for determining this value is given by

\begin{equation}
    \label{eq:spectral_gap}
    \begin{aligned}
    \quad & \mathscr{L}_{norm}\nu = \lambda\nu\\
    \textrm{s.t.} \quad & \det(\mathscr{L}_{norm}-\lambda\mathscr{I}) = 0
    \end{aligned}
\end{equation} 
where, $\mathscr{L}_{norm}$ is the normalized Laplacian matrix, $\nu$ is the eigenvector, $\lambda$ is the eigenvalue, $det(.)$ represents the determinant and $\mathscr{I}$ is the identity matrix of the same size as $\mathscr{L}_{norm}$. Solving the equation yields $\lambda_0, \lambda_1,...,\lambda_{n-1}$ and $\lambda_1$ is the first non trivial eigenvalue which is the spectral gap. In general, this paper considers 1) the star network, for simple control that can tolerate a single point of failure; and 2) the radial network, for a structured hierarchy that has some redundancy. 
These two network topologies cover the most common communication structures for power transmission networks. Typically, power grid communication networks are sparse, as the infrastructure is designed to minimize connections and maintain efficiency, making expander network unrealistic in cyber-physical power system infrastructure networks. However, for certain flat cyber network configurations that exhibit high connectivity (e.g., communication paths allowed by firewall rules), our proposed method can be generalized to these cyber network configurations. Broadly, by adopting the developed DESTinE framework, one can generalize it to model any kind of communication networks for power systems. 
%

\textbf{Global Clustering Coefficient}: This coefficient measures the degree to which the neighbors of a given vertex are interconnected. While the local clustering coefficient applies to individual vertices, the global clustering coefficient is derived by averaging the local coefficients across all vertices, resulting in a value between 0 and 1. This global measure provides insight into the overall interconnectedness of the network \cite{aksoy2017measuring}.

The results of the global matrix analysis for the power system test cases are presented in Table \ref{tab: global matrix}, offering a quantitative evaluation of the network's structural properties, including its potential for subdivision and clustering. The analysis indicates that for both star and radial topologies, each node exhibits sparse connectivity, as reflected by values that are either zero or very close to zero. This observation holds consistently across all three cases, highlighting the limited inter-node connectivity within the network. Consequently, the network can be effectively partitioned into subgraphs. The sparse connectivity and inherent partitionability of these graphs ensure that the optimization process described in Section \ref{result} achieves faster convergence with guaranteed reliability.


\subsection{Local Matrix}
We assess three key node centrality measures, i.e., betweenness centrality, eigenvector centrality, and closeness centrality  due to their effectiveness in identifying critical nodes and evaluating network robustness \cite{borgatti2006graph, bonacich2007some}, defined as 



\textbf{Betweenness Centrality}: It determines the significance of a node within a network by evaluating the node’s central position. Betweenness centrality measures how frequently a node serves as a connecting point along the shortest path between two other nodes. The betweenness centrality is calculated by  
%
\begin{equation}
C_B(v) = \sum_{s \neq v \neq t \in V} \frac{\sigma_{st}(v)}{\sigma_{st}}
\label{eq:bet_cen}
\end{equation}
where $C_B(v)$ is the betweenness centrality of node $v$, $\sigma_{st}$ is the total number of shortest paths from node $s$ to node $t$, and $\sigma_{st}(v)$ is the number of those paths that pass through node $v$.

\textbf{Eigenvector Centrality}: It is a network centrality measure that assigns relative scores to all nodes, based on the principle that connections to highly scored nodes contribute more to a node’s score than connections to lower-scoring nodes. It measures the influence or importance of a node by considering both direct and indirect connections. The eigenvector centrality $X(v)$ of node $v$ in the network $G$ is calculated by

\begin{equation}
\label{eq:eigenvector}
X(v)=\frac{1}{\lambda}\sum_{u\in N(v)}X(u)
\end{equation}
where $N(v)$ is the set of neighbors of node $v$, $X(u)$ is the centrality score of the neighbor node $u$, and $\lambda$ is the dominant eigenvalue of the adjacency matrix of the network.

\textbf{Closeness Centrality}: It is the inverse of the sum of the shortest path distances from a given node to all other nodes in the graph. Mathematically, it can be represented by

\begin{equation}
    \label{eq:close_cen}
    C_C(v) = \frac{1}{\sum_{u\in V}s(v,u)}
\end{equation}
where $s(v, u)$ is the shortest path distance between node $v$ and node $u$, with the sum taken over all nodes u in the graph. In essence, closeness centrality quantifies how quickly a node can reach every other node in the network. A higher value of $C_C(v)$ indicates that the node is more central, having shorter average paths to other nodes, thus making it efficient for transmitting information or resources. This metric is particularly valuable for identifying key nodes in networks where distance or reachability is critical.

\subsection{Optimization with Local Matrix}
All three cases described previously are used to establish the ranking of critical elements by combining the local matrix with simulation results. 
An objective function is introduced to assign specific weights to both the local matrix and the simulation results, with a particular focus on ranking critical nodes, enabling a balanced and prioritized assessment of critical network elements.
The analysis specifically targets utilities as critical nodes, and given the subgraph structure, this approach is transferable to other similar networks. It can also be extended to generate rankings of other critical elements, such as substations and regulatory units, ensuring broader applicability across different network configurations.


Using the local matrix, we rank the critical utilities using $\mathbb{C}_{B}(v)$, $\mathbb{X}(v)$, and $\mathbb{C}_C(v)$ that represent the rankings of critical utilities based on betweenness centrality, eigenvector centrality, and closeness centrality matrices, respectively. We further formulate an objective function to determine an alternative ranking of critical utilities, denoted as $T(v)$, which incorporates the ranking obtained from discrete-event simulation,  $S(v)$. The optimization problem, including the objective function and the constraints, is given by 
%
\begin{equation}
    \label{eq:optimization_1}
    \begin{aligned}
    \min_{a,b,c,T(v)} \quad & \frac{1}{a} {\left| T(v)- \mathbb{C}_B(v) \right|}^2 + \frac{1}{b} {\left| T(v)- \mathbb{X}(v) \right|}^2 + \\
    & \frac{1}{c} {\left| T(v)- \mathbb{C}_c(v) \right|}^2\\
    \textrm{s.t.} \quad & 0\le \left| S(v)- T(v) \right| \le 1\\
     & a + b + c = 1\\
     & a, b, c  > 0 \\
     & 0 \le T(v), S(v), \mathbb{C}_B(v), \mathbb{X}(v), \mathbb{C}_c(v) \le k\\
    \end{aligned}
\end{equation} 
where $a$, $b$, $c$, and $T(v)$ are the decision variables. The matrices $\mathbb{C}_B(v)$, $\mathbb{X}(v)$, $\mathbb{C}_C(v)$, $S(v)$, and the decision variable $T(v)$ 
consists of integer values ranging from 0 to $k$, where $k$ varies depending on the test case. Specifically, $k$ is 3 for the ACTIVSg500 case, 19 for the ACTIVSg2000 case, and 79 for the ACTIVSg10k case, corresponding to the 4, 20, and 80 utilities within the respective network configurations. The constraint $0 \le \left| S(v)- T(v) \right| \le 1$  ensures that the rank of the elements either matches or differs by at most one. While minimizing the objective function determines $T(v)$, it is observed that $T(v)$ is heavily influenced by $S(v)$. To mitigate this dependency, we apply bounds, $a, b, c  > 0$ and $a + b + c = 1$,   to the decision variables. 
We incorporate a penalty term into the revised objective function, designed to reduce the influence of $S(v)$ on $T(v)$. Further elaboration on this improvement is given in Section \ref{result}.





\section{Results in Normal and Adversarial Condition with Optimization}
\label{result}

\begin{figure}[!t]
    \centering
    \begin{subfigure}{0.45\textwidth}
         \centering
         \includegraphics[width=\textwidth]{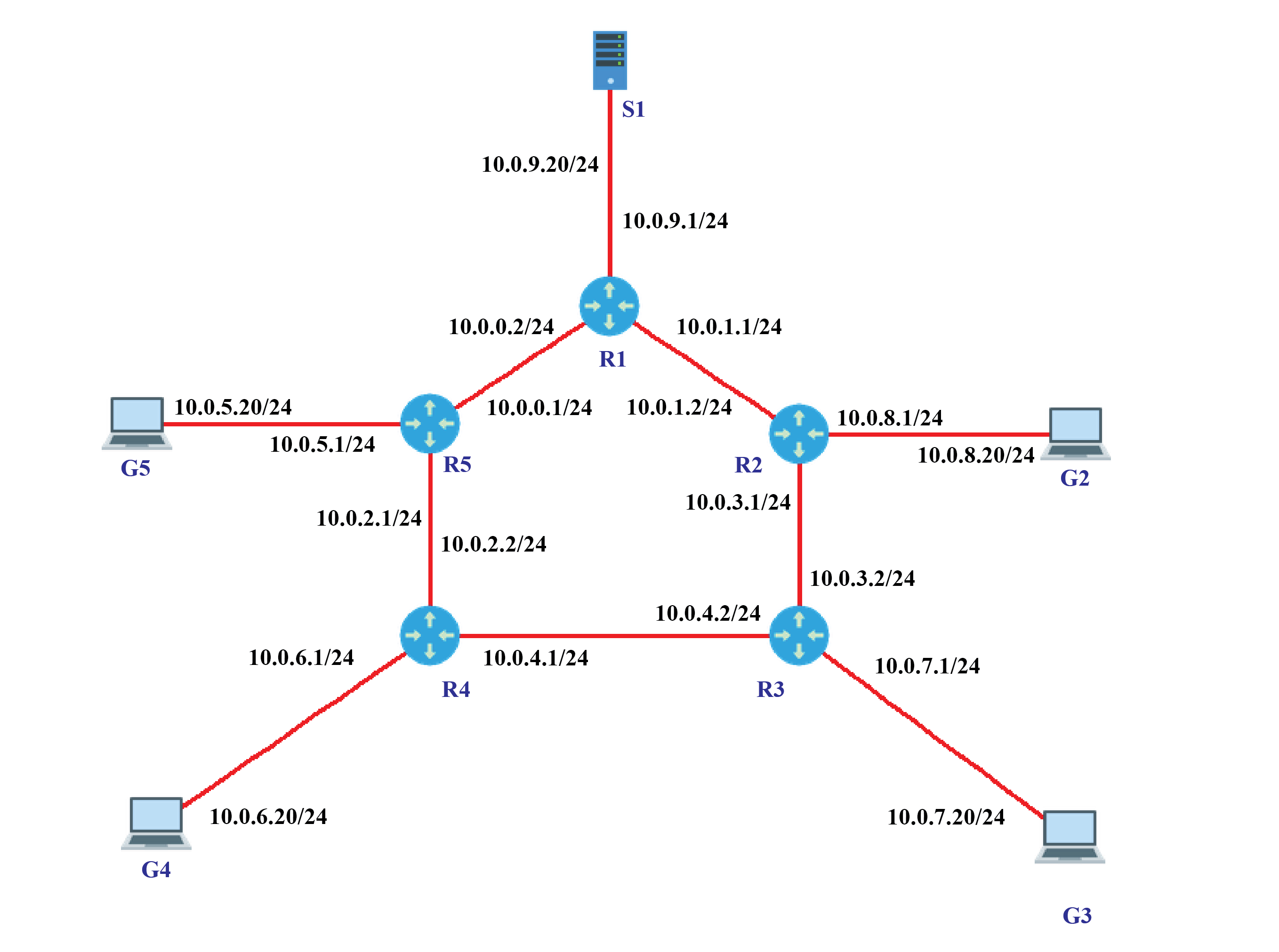}
         \caption{}
     \end{subfigure}
     
     \begin{subfigure}{0.37\textwidth}
         \centering
         \includegraphics[width=\textwidth]{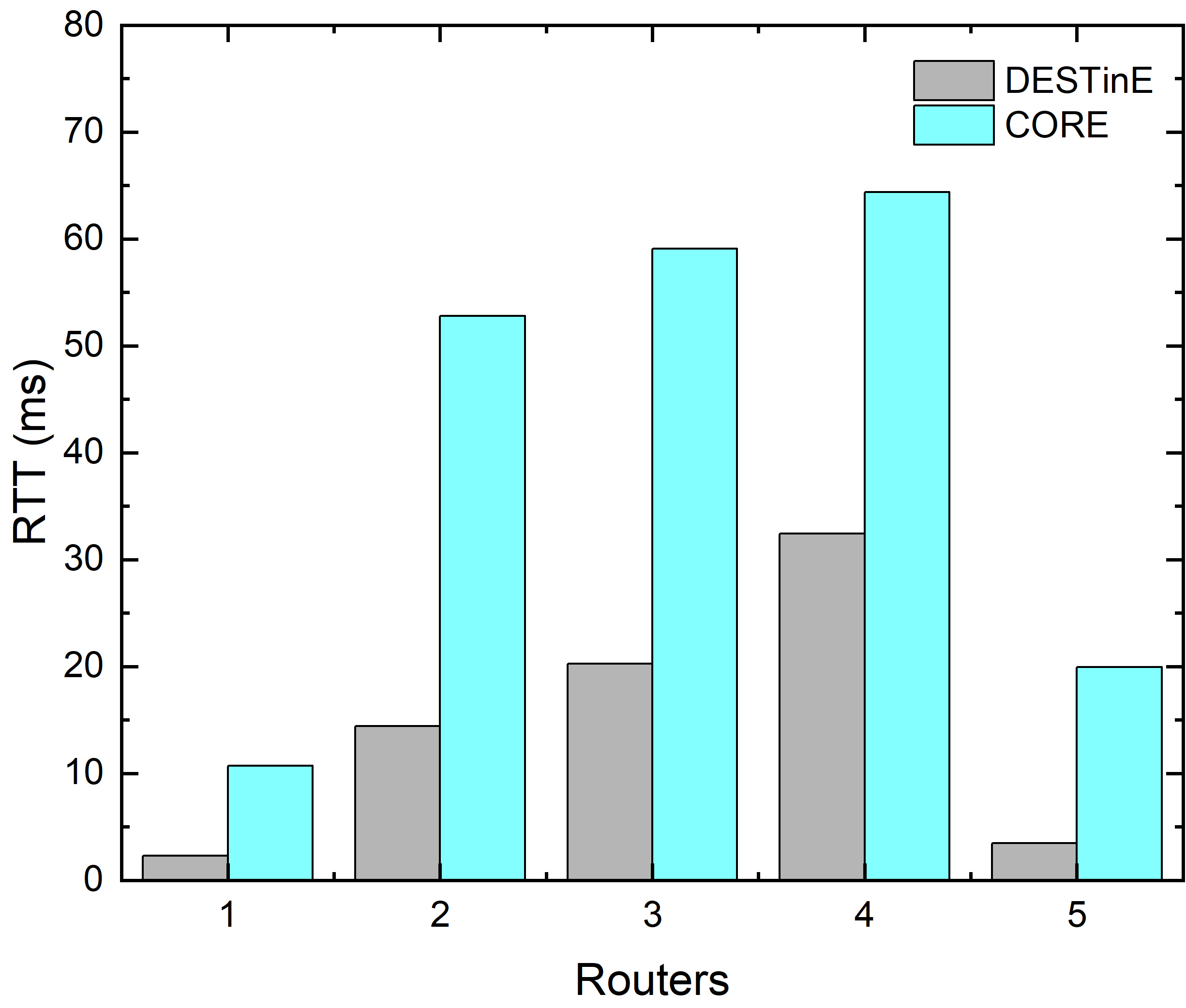}
         \caption{}
     \end{subfigure}
     \captionsetup{justification=justified}
    \caption{Comparison of the simulation tool with Common Open Research Emulator (CORE) showing (a) network topology and (b) results for normal condition. }
    \label{fig:core}
\end{figure}


\begin{figure}[!t]
    \centering
    \begin{subfigure}{0.45\textwidth}
         \centering
         \includegraphics[width=\textwidth]{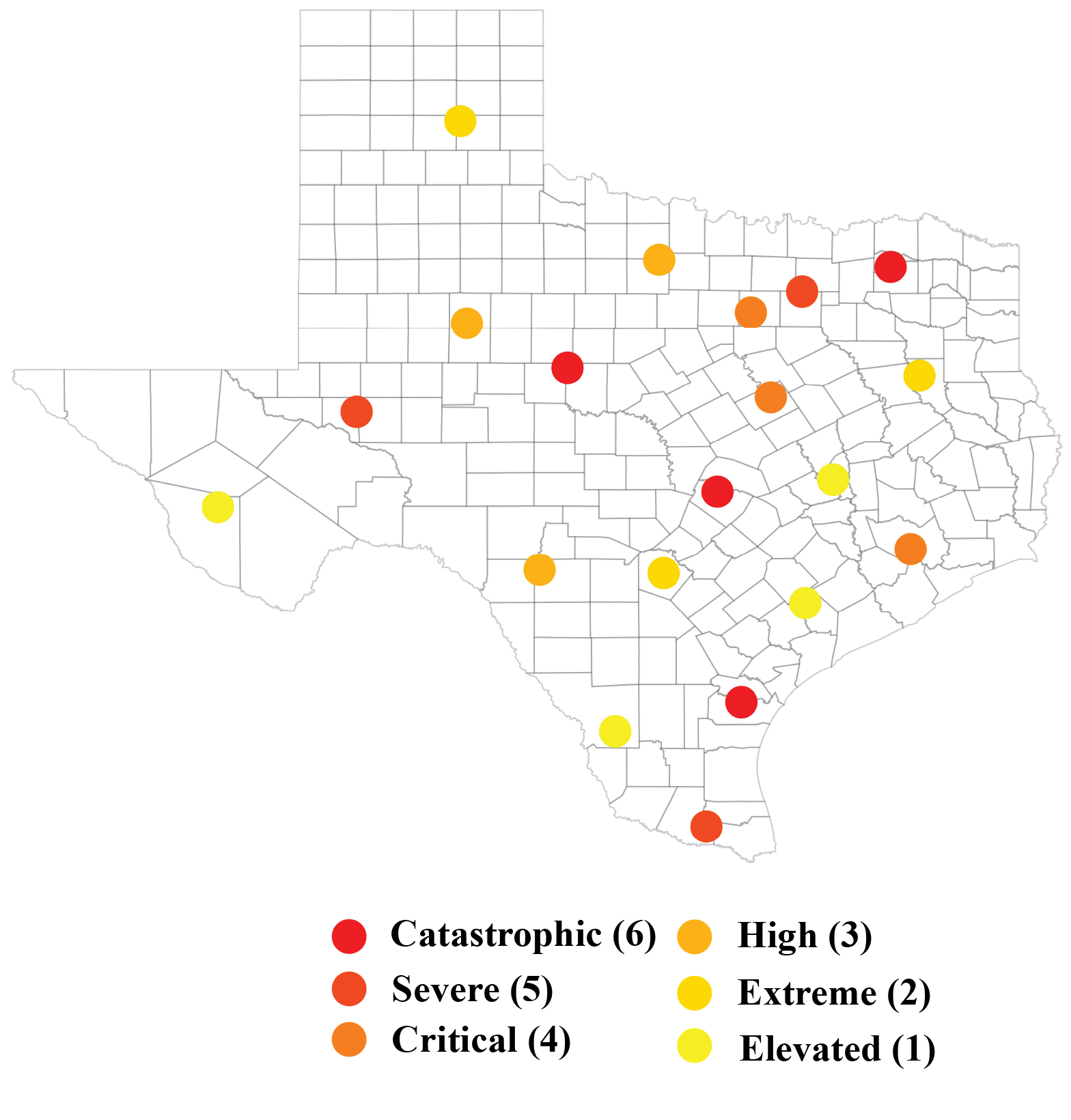}
         \caption{}
     \end{subfigure}\hfill
     
     \begin{subfigure}{0.45\textwidth}
         \centering
         \includegraphics[width=\textwidth]{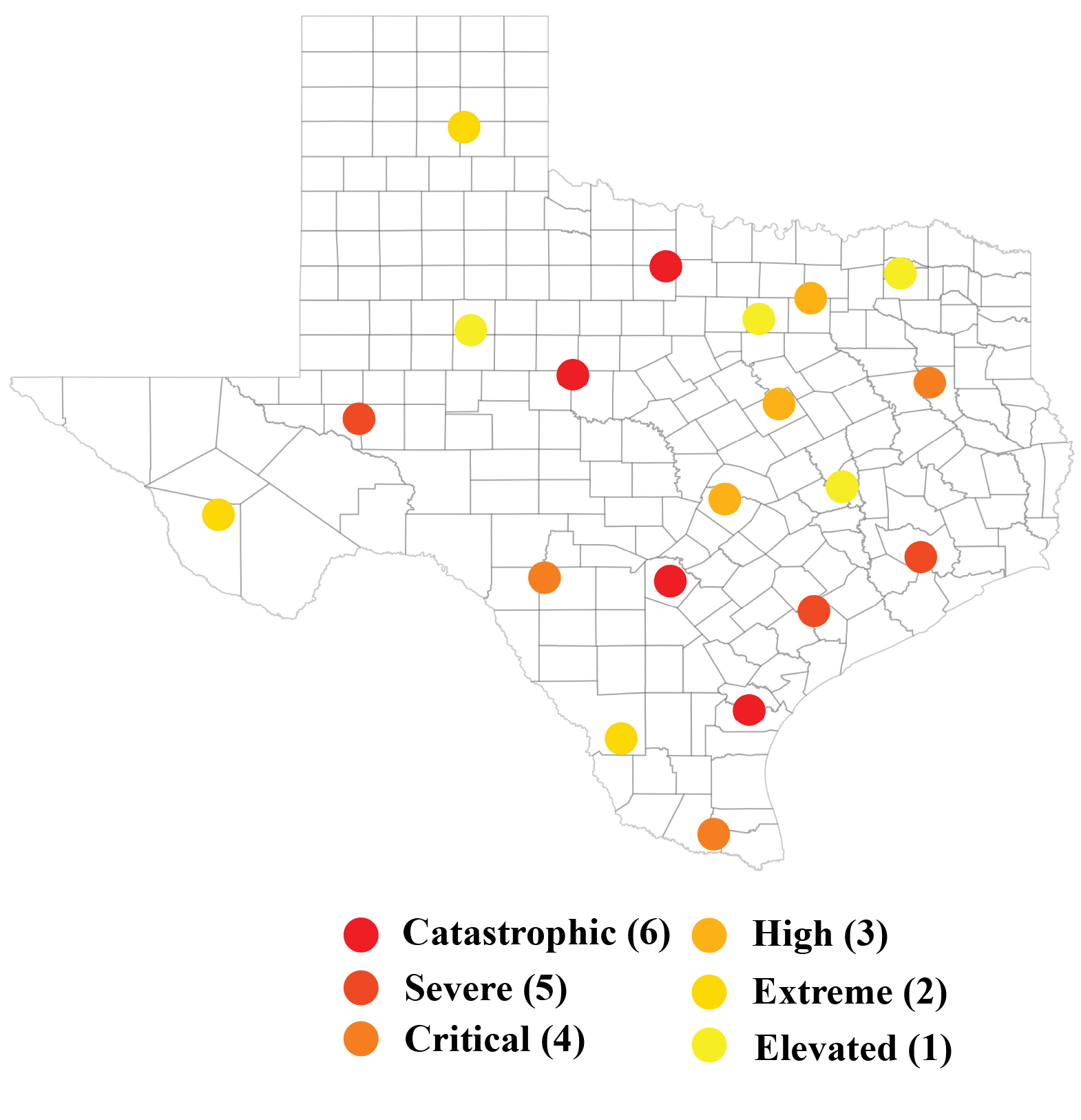}
         \caption{}
     \end{subfigure}
     \captionsetup{justification=justified}
    \caption{Severity classifications of DoS impact on utility for ACTIVSg2000 case where each circle represents a utility and the color of the circle represents the classification for (a) star topology and (b) radial topology. }
    \label{fig:dos_2kk}
\end{figure}

\begin{figure}[!t]
    \centering
    \begin{subfigure}{0.4\textwidth}
         \centering
         \includegraphics[width=\textwidth]{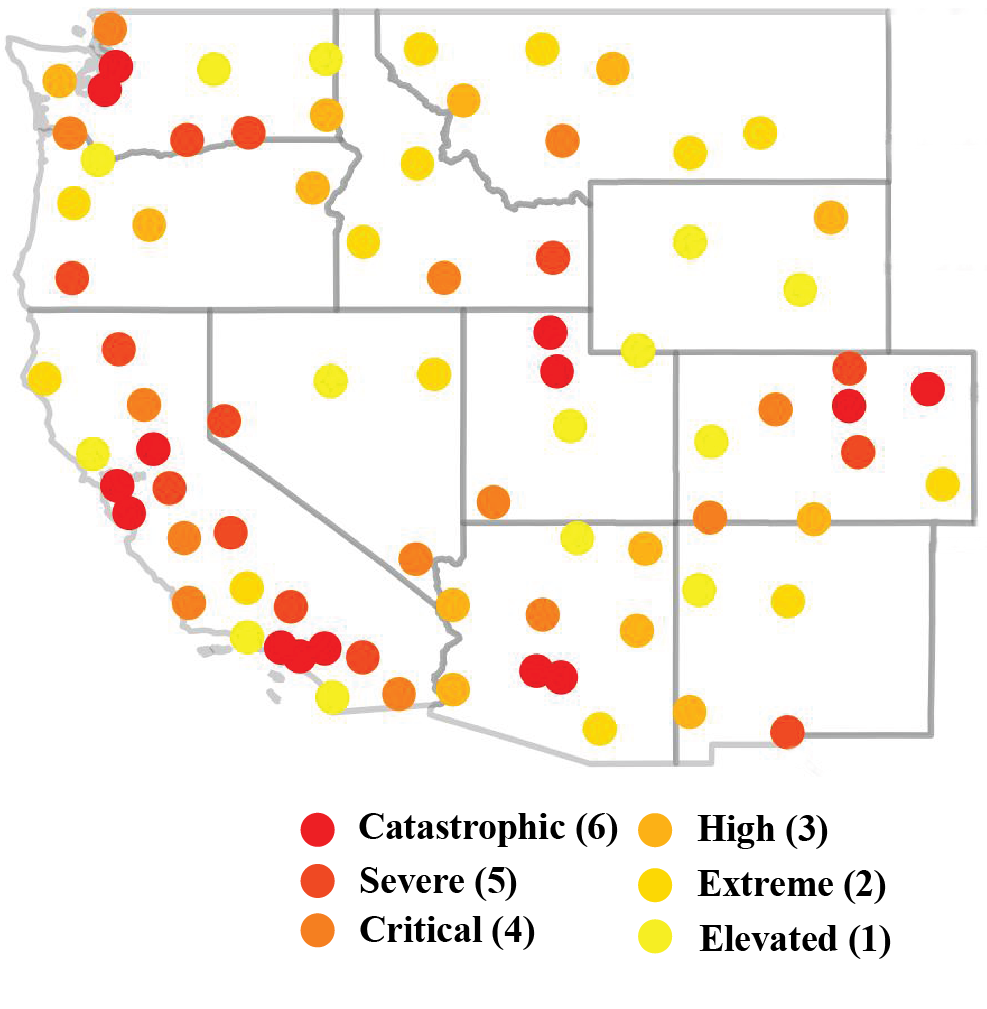}
         \caption{}
     \end{subfigure}\hfill
     
     \begin{subfigure}{0.42\textwidth}
         \centering
         \includegraphics[width=\textwidth]{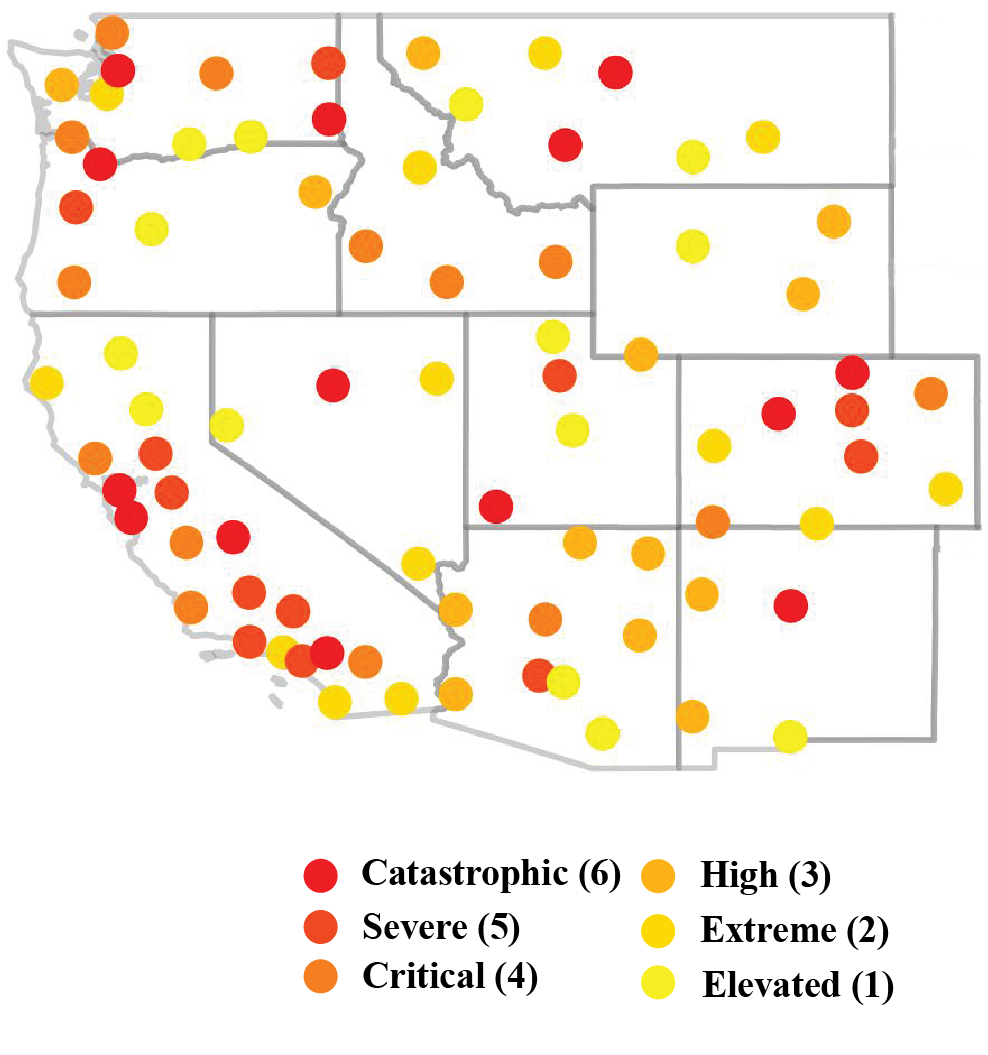}
         \caption{}
     \end{subfigure}
     \captionsetup{justification=justified}
    \caption{Severity classifications of DoS impact on utility for ACTIVSg10k case where each circle represents a utility and the color of the circle represents the classification (a) Star topology and (b) Radial topology. }
    \label{fig:dos_10k}
\end{figure}

Next, we present the results under both normal and adversarial conditions.
%
The pseudo-code and the complete result tables for all cases are given in the \nameref{appendix}. The pseudo-code is divided into smaller, more manageable sections to enhance readability. Pseudo-codes \ref{algo1}-\ref{algo3} are for the star network configuration, and Pseudo-codes \ref{algo4}-\ref{algo7} are tailored for the radial network configuration.

\subsection{Normal Condition}
In this setup, we utilize the discrete-event simulator to model power system operations for a duration of 1,000-time units, equivalent to 1,000 seconds in real time. The simulation was conducted on three distinct power system cases, each evaluated under both star and radial topologies. While the simulated time span corresponded to 1,000 seconds, the simulator executed much faster than real-time, providing rapid insights and avoiding the extended computational delays often associated with traditional simulators.

As an example, we consider a scenario where the system experiences a single-source Denial-of-Service (DoS) attack. In this case, an adversary floods a key communication channel with excessive data packets, overwhelming the bandwidth and delaying critical control messages, such as those sent to SCADA systems or protection relays. Using the discrete-event simulator, we model the progression of the attack, the resulting communication bottleneck, and its impact on the system. By running the simulation in a fraction of real-time, we 
%
are able to assess the system's vulnerabilities to such attacks and test mitigation strategies, such as prioritizing command traffic or implementing redundancy in communication channels. 



As detailed in Table \ref{tab1}, each entity acts as a router, facilitating packet forwarding from the packet generator to the sink. The packet generator connects to a substation, with packets collected at the utility and the regulatory unit sinks. Our primary metric focuses on the average delay for routers at the utility and regulatory units, whereas substations could also be included. Notably, both simulation and local matrix analysis identify utility nodes as the most critical, supporting the focus on these elements. 



In the ACTIVSg500-bus and ACTIVSg2000-bus power system cases, there is only one regulatory unit, which acts as a single point of failure. Consequently, regulatory nodes are excluded under adversarial conditions. Tables \ref{tab:delay_normal_500}, \ref{tab:delay_normal_2k} and \ref{tab:delay_normal_west_us} in the \nameref{appendix} provide a ranking of the utility routers for all test cases, along with their corresponding rankings derived from the local matrices.


\subsection{Adversarial Condition}

In this section, we launched a DoS attack to resemble adversarial disturbances, while other threat models are also applicable. 
The DoS attack is applied to all test cases (ACTIVSg500, ACTIVSg2000, and ACTIVSg10k), including both the star and radial topology. Specifically, we targeted each of the utility routers in the network.

To simulate the DoS attack, we significantly reduced the packet forwarding probability of each router to a minimal level. Under normal conditions, routers forward packets with a specified probability, but in this scenario, that probability was drastically lowered, resulting in most packets being dropped rather than forwarded. This effectively transforms the router into a bottleneck, leading to substantial packet loss and degraded network performance, which mimics a real-world DoS attack.
Within the SimPy environment, the forwarding probability for each affected router was set to 0.01 to simulate the DoS attack. The simulation was then run for a predefined duration, allowing us to evaluate the network's behavior under these adversarial conditions. This approach was specifically chosen to simulate network degradation in a controlled manner, by reducing the packet forwarding probability instead of introducing attacker node packet generators, which could result in complete packet loss and render the simulation ineffective. By doing so, the simulation ensures minimal packet loss while allowing for a measurable and realistic impact on the network performance. This method provides valuable insights into the effects of a DoS attack, particularly in terms of increased delay, packet loss, and the creation of bottlenecks at critical routers.

While DoS is chosen as a representative attack scenario in this study, the DESTinE framework is designed to flexibly simulate a range of cyberattack models within its software environment. Attacks such as false data injection (FDIA), man-in-the-middle (MITM), replay attacks, and side-channel attacks can be synthesized by modifying packet content, flow dynamics, routing behavior, or event triggers within DESTinE’s event-driven simulation loop—without the need for hardware integration.

\begin{itemize}
    \item FDIA can be simulated by dynamically altering packet data during transmission.
    \item MITM attacks can be modeled by inserting a malicious node that intercepts and manipulates packet content.
    \item Replay attacks can be executed by capturing transmitted packets and re-injecting them into the network.
\end{itemize}

These attack templates can be seamlessly integrated into DESTinE’s architecture, allowing the platform to simulate a wide range of complex adversarial scenarios purely in software-only mode. This flexibility ensures DESTinE’s capability to support future research and analysis on diverse cyber threats impacting cyber-physical energy systems.

\subsection{Comparison of \name with the Common Open Research Emulator (CORE)}

The comparison was set up using a modified small network, as described in \cite{haque2024graphtheoryvstimedomain}. This network consists of five routers, with four routers connected to packet generators and one router linked to the sink in accordance with our simulation tool. We also model the same topology in the Common Open Research Emulator (CORE). In this configuration, the routers are labeled R1 to R5, the packet generators are labeled G2 to G5, and the sink is labeled S1, as shown in Fig. \ref{fig:core}. To facilitate an equal comparison, we developed a bash script that ensures each packet generator transmits the same number of packets in our simulation tool \name and CORE. Additionally, the packet sizes follow an exponential distribution with a mean size of 100 bytes, and packets are sent at intervals of 0.5 seconds, also adhering to an exponential distribution. The packets are transmitted using the User Datagram Protocol (UDP), which accommodates the transmission of variable-sized packets.

While our implementation does not replicate the simulation tool exactly; rather, we aimed to create a comparable environment to minimize disparities between our simulation tool and the CORE emulator. In CORE, all packets are captured as a packet capture file, enabling the calculation of the round-trip time (RTT) for each packet and deriving the average RTT, shown on the $y$-axis in Fig. \ref{fig:core} (b). These average RTTs for each router provide an indication of the average delay observed in our simulation tool. While absolute values differ, the relative delay order from our simulation tool aligns closely with CORE's. Importantly, \name offers superior speed and scalability, enabling efficient analysis of large network models. In contrast, CORE's resource-intensive processes require up to 1000 seconds to emulate 400 nodes \cite{tan2012score}, primarily due to the overhead of generating individual artifacts such as virtual routers, packet generators, and sinks for each node. This significantly limits its scalability for large network simulations. On the other hand, DESTinE achieves significantly faster execution, especially when integrated with HIL systems. In our experiments, DESTinE successfully simulated over 4700 nodes in approximately 200 seconds with HIL integration, demonstrating its efficiency and scalability for large-scale network simulations.  A detailed analysis incorporating HIL integration is provided later in Section \ref{vser_hil_destine} of  the manuscript to further demonstrate DESTinE’s efficiency in large-scale network simulations.



\subsection{Optimized Result for Stable and Adversarial Condition}
In our analysis of the three test cases, we evaluated the rank of utility routers under both star and radial topologies, as well as the ranks based on local matrix measures (i.e., betweenness, eigenvector, and closeness centrality). During the DoS disturbance, we recalculated the rank of each utility router by removing the affected nodes from the network and recalculating the centrality matrices. These recalculated values were then used to re-rank the utility routers, providing a clearer view of how the attack impacts the network's critical nodes.

To optimize the ranking process, we modified the objective function (Equation \eqref{eq:optimization_1}) to ensure it remains a convex optimization problem. A slack variable, $\alpha$, was introduced to act as a penalty term for any deviation from the imposed constraints. Moreover, for the constraint formulation, we chose to use the square of the difference instead of the absolute difference, as this adjustment produced better results. Squaring the difference enhances the smoothness of the optimization process, reduces potential discrepancies, and provides a more stable solution. This method also serves to relax the dependency on the rank obtained from the simulation, allowing for a more flexible and optimized approach to ranking the utility routers. Additionally, some constraints were relaxed to improve flexibility in the optimization process. Our analysis revealed that the simulation results are highly correlated with the betweenness centrality matrix, followed by the eigenvector and closeness centrality matrices, which guided us to place bounds on these decision variables for more robust results.


We derived two critical values for $\alpha$: $\alpha_0$ and $\alpha_1$. $\alpha_1$ corresponds to the scenario where the rank of each utility router from the simulation results differs by 1 from the rank obtained via the optimized matrix. In contrast, $\alpha_0$ represents the case where the simulation rank perfectly matches the optimized matrix rank, with no difference. This process ensures that the optimization is both precise and efficient. The modified objective function, incorporating these constraints and bounds, is presented in Equation \eqref{eq:optimization_slack}. This draws a parallel to the $(N-1)$ contingency analysis used in Security-Constrained Optimal Power Flow (SCOPF)~\cite{wood2013power}.
\name achieves this by quantitatively assessing the impact of sequential router removal through DoS threat and observing the effects on overall network performance. Thus, \name
is well-suited for network contingency analysis and optimization.
%



\begin{equation}
    \label{eq:optimization_slack}
    \begin{aligned}
    \min_{a,b,c,T_{DoS}(v)} \quad & \frac{1}{a} {\left| T_{DoS}(v)- \mathbb{C}_{B_{DoS}(v)} \right|}^2 + \\
    & \frac{1}{b} {\left| T_{DoS}(v)- \mathbb{X}_{DoS}(v) \right|}^2 + \\
    & \frac{1}{c} {\left| T_{DoS}(v)- \mathbb{C}_{c_{DoS}(v)} \right|}^2 +\\
    & \alpha{\left| S_{DoS}(v)- T_{DoS}(v) \right|}^2\\
    \textrm{s.t.} \quad & c \le b \le a\\
    & a, b, c > 0\\
    & a + b + c = 1\\
    & 0 \le T_{DoS}(v), S_{DoS}(v), \mathbb{C}_{B_{DoS}(v)},  \mathbb{X}_{DoS}(v),\\ 
    & \mathbb{C}_{c_{DoS}(v)} \le k-1\\
    & {\left| S_{DoS}(v)- T_{DoS}(v) \right|}^2=0  \quad (\alpha = \alpha_0)\\
    & {\left| S_{DoS}(v)- T_{DoS}(v) \right|}^2=1  \quad (\alpha = \alpha_1)\\
    \end{aligned}
\end{equation}


\begin{table*}
\centering
\captionsetup{justification=justified}
\caption{Simulation Completion Times for DESTinE Across Different Configurations, Topologies, and ACTIVSg Cases (NB: The number of packets generated and collected for each case is mentioned in parentheses.)}
\label{tab:destine_hil}
\begin{tblr}{
  column{3} = {c},
  cell{1}{1} = {r=2}{},
  cell{1}{2} = {c=2}{},
  cell{1}{4} = {c=3}{},
  cell{1}{7} = {c=3}{},
  vlines,
  hline{1,3-6} = {-}{},
  hline{2} = {2-9}{},
}
\textbf{ Configuration } & \textbf{ ACTIVSg500 (481 Packets) } &  & \textbf{ ACTIVSg2000 (2449 Packets) } &  &  & \textbf{ ACTIVSg10k (5707 Packets) } &  & \\
 & \textbf{ Star } & \textbf{ Radial } & \textbf{ Star } & \textbf{ Radial } & \textbf{Hybrid } & \textbf{ Star  } & \textbf{ Radial } & \textbf{Hybrid }\\
DESTinE & 0.09s & 0.08s & 0.51s & 0.48s & 0.44s & 1.45s & 1.44s & 1.42s\\
DESTinE + Virtual Server & 1.86s & 1.56s & 2.64s & 1.91s & 1.86s  & 6.64s & 4.00s & 3.91s\\
DESTinE + Raspberry Pi 5 & 17.00s & 16.42s & 89.00s & 87.45s & 88.44 s & 208.00 s & 209.10s & 208.00 s
\end{tblr}
\end{table*}

In \eqref{eq:optimization_slack}, the subscript $DoS$ indicates that the decision variables are evaluated after a DoS attack is initiated, and the specific element in question has been excluded from this calculation.  For the last two constraints in problem \eqref{eq:optimization_slack}, $\alpha_1$ represents a one-rank difference between the simulation and optimized matrix rankings, while $\alpha_0$ indicates a perfect match between the simulation rank and the optimized matrix rank. A higher value of $\alpha_0$ and $\alpha_1$ indicates a stronger reliance on the simulation-based ranking of the utility routers, emphasizing the importance of this ranking in determining network performance. The minimization problem yields consistent values for parameters $a$, $b$, and $c$ under both normal and adversarial conditions, which are set to be 0.97, 0.02, and 0.01, respectively. These values correspond to the weights assigned in the centrality matrices.  In contrast, the values of $\alpha_0$ and $\alpha_1$ can be equivalent to the order of $a$, $b$, and $c$ or scaled in multiples of 100 relative to $a$, $b$, and $c$. These values vary based on the network conditions in the three test cases. A DoS disturbance is an unwanted event that can severely disrupt the network. In our simulation, a DoS attack can incapacitate several substations and affect a regulatory unit. By analyzing the combined effects of substations taken offline and the relative increases in $\alpha_0$ and $\alpha_1$ values compared to the network's normal state, we developed a classification system to assess the criticality of a utility router under a DoS attack. In this system, greater weight is assigned to $\alpha_0$, followed by the number of substations lost, and then $\alpha_1$. 
However, it is important to note that the ACTIVSg500 bus system case consists of only four utilities. As such, this classification system is not applicable to this test case. Should the network topology support a larger number of utilities, these nodes could be classified accordingly.
The classifications are ranked from most to least severe as follows: Catastrophic (6), Severe (5), Critical (4), High (3), Extreme (2), and Elevated (1). The numerical values in parentheses represent the degree of compromise of an element, with higher numbers indicating greater levels of compromise. A `Normal' classification signifies that no DoS attack is present in the network.

In the Appendix, Tables \ref{tab:dos_500}, \ref{tab:dos_2k_star}, \ref{tab:dos_2k_radial}, \ref{tab:dos_star}, and \ref{tab:dos_radial} provide a detailed classification of utility routers for the three cases, including the corresponding $\alpha_0$ and $\alpha_1$ values 
for star and radial topologies. These values guide the reconfiguration of the network, illustrating which 
utilities are more susceptible or resilient under different conditions. The decreasing (or increasing) of $\alpha_0$ and $\alpha_1$ depend on the network structures of the power system case. For example, in the ACTIVSg2000 case, a bigger $\alpha_0$ corresponds to the loss of a greater number of substations, which can be perceived as leading to increased power loss.. Subsequently, in both Table \ref{tab:dos_2k_star} and Table  \ref{tab:dos_2k_radial}  (DoS threat in ACTIVSg2000) and Table \ref{tab:dos_star} and Table \ref{tab:dos_radial}  (DoS threat in ACTIVSg10k), the values of both $\alpha_0$ and $\alpha_1$ present a decreasing trend when the severity level decreases from Catastrophic (6) to Elevated (1). 
However, we note that for the DoS threat in the ACTIVSg10k case, the values of $\alpha_0$ and $\alpha_1$ are not exactly monotonically decreasing, but with an overall decreasing trend when fewer substations are affected by the DoS attack. For clarity, Figs. 
\ref{fig:dos_2kk} and \ref{fig:dos_10k} present a visual representation of the severity classifications of DoS impact on the 500 and 10,000-bus systems. This holistic approach aids in decision-making, allowing network operators to mitigate vulnerabilities based on topological insights. 


The severity classifications of DoS impacts are influenced not only by the nature of the attack but also by the network topology, such as whether it is arranged in a star or radial configuration. With \name
that operates at speeds faster than real-time, network designers can perform detailed, real-time analyses of these impacts. Additionally, the tool is capable of simulating a variety of network configurations beyond star and radial topologies, offering significant flexibility in network analysis. For example, if an element is identified as less critical in a star topology, the network section can be strategically reconfigured from radial to star, or vice versa, to bolster resilience against adversarial conditions. This dynamic capability enhances the network's defense against DoS attacks by optimizing the placement and organization of critical nodes.






\subsection{Risk Assessment and Hybrid Topology Proposal}
After quantifying the risk of utilities for the ACTiVSG2000 and ACTiVSG10k bus systems, a hybrid topology is proposed to enhance system efficiency. This hybrid topology is formulated by modifying the existing structure based on a comparative risk analysis of star and radial topologies. The modification process follows a systematic approach to selecting the optimal utility connection for each substation, ensuring both risk minimization and efficient packet transmission.

The DESTinE optimizer provides a numerical severity of impact value ranging from 1 to 6, where 1 represents the lowest risk and 6 represents the highest risk for each substation under the star and radial topologies. The selection of the hybrid topology is guided by the following criteria:
\begin{itemize}
    \item If the severity of impact in the star topology is lower than in the radial topology, the star topology is selected.
    \item If the severity of impact in the radial topology is lower than in the star topology, the radial topology is selected.
    \item If the severity of impact values are equal for both topologies, the radial topology is chosen by default, as it generally provides better performance in packet transmission.
\end{itemize}
Mathematically, the topology selection for a given substation sss can be defined as follows:


\begin{equation}
\label{eq:hybrid_topology}    
T_h(s) =
    \begin{cases}  
        \text{Star}, & \text{if } R_s < R_r \\  
        \text{Radial}, & \text{if } R_r < R_s \\  
        \text{Radial}, & \text{if } R_s = R_r  
    \end{cases} 
    \quad  
    \begin{aligned}
        &\forall s \in S, \\  
        &R_s, R_r \in \{1,2,3,4,5,6\}
    \end{aligned}
\end{equation}

where, $S$ is the set of all  substations; $R_s$ and $R_r$ denote the severity of impact values (ranging from 1 to 6) for star and radial topologies, respectively, as provided by the DESTinE optimizer; and $T_h(s)$ represents the topology assignment function for substation $s$.\\

The hybrid topology is formed by iterating through all substations and applying $T_h(s)$ to determine the optimal connection type. This approach ensures a risk-aware network configuration that leverages the benefits of both topologies while minimizing system vulnerabilities.

\subsection{Performance Evaluation of DESTinE with Virtual Server and Hardware Integration}
\label{vser_hil_destine}

The performance of the DESTinE simulator was evaluated under controlled conditions by generating packets of uniform size and quantity across three test cases, each utilizing a comparable network topology. The assessment involved two configurations: one where DESTinE was connected to a Node.js-based virtual server and another where it was integrated with a Raspberry Pi 5 device. Implemented in Python, DESTinE operated within these environments, while the virtual server functioned as an external computational platform to simulate a distributed network.

The Raspberry Pi 5 has an estimated maximum throughput of approximately 200 MBPS \cite{rpi5200mbps}. However, since multiple applications were running concurrently, the router throughput was capped at 60 MBPS to maintain a stable and optimized connection between DESTinE and the Raspberry Pi 5. It is important to note that this ensures a realistic implementation, reflecting practical network constraints that might be encountered in real-world deployments.

The test case was structured such that each router could generate at least one packet but no more than two packets, with packets being generated at 50 ms intervals. The simulation was run for a sufficient duration to allow all packets originating from substation routers to reach the regulatory unit’s router. Three configurations were evaluated:

\begin{enumerate}
    \item Standalone DESTinE: The entire simulation was executed within DESTinE, and the actual completion time of the simulation was recorded.
    \item DESTinE with a Virtual Server: The substation routers were implemented inside DESTinE, while the utilities and regulatory unit routers were hosted on a Node.js-based virtual server. Packets generated from the substation routers within DESTinE were transmitted to the virtual server, where they were collected and processed. The real-time duration required for this data exchange was measured.
    \item DESTinE with a Raspberry Pi 5: DESTinE was connected to a Raspberry Pi 5 over a WiFi 6 network. The substation routers were implemented within DESTinE, while the utilities and regulatory unit routers were hosted inside the Raspberry Pi. The Raspberry Pi’s virtual ports were used to represent the routers of these units.
\end{enumerate}
It was observed that the stability of the connection between DESTinE and the Raspberry Pi depended on the number of routers in the utilities and regulatory units.
\begin{itemize}
    \item If the total number of these routers was fewer than 25, stable connections were established using ports starting from 5000 and 6000.
    \item If the total number exceeded 25, ports starting from 35,000 or 36,000 provided improved stability.
\end{itemize}

The simulation completion times for all three configurations are presented in Table \ref{tab:destine_hil}, where an equal number of packets of uniform size were generated and collected for each topology. The results indicate that, in both standalone DESTinE and DESTinE with a virtual server, the star topology required the longest time, followed by the radial topology. At the same time, the proposed hybrid model achieved the shortest completion time.

However, a more nuanced outcome was observed when DESTinE was connected to the Raspberry Pi 5, particularly for the ACTIVSg2000 and ACTIVSg10k cases.
\begin{itemize}
    \item For the ACTIVSg2000 case, the star topology required the longest time, followed by the proposed hybrid topology, while the radial topology exhibited the shortest completion time.
    \item For the ACTIVSg10k case, the star and hybrid topologies required equal completion times, whereas the radial topology took the longest time.
\end{itemize}

DESTinE efficiently represents network structures using graph-based data structures, specifically an adjacency list, to facilitate packet routing and risk evaluation while supporting topology modifications for seamless integration of hybrid models based on real-time risk assessments.
Due to the large scale of the test cases, the figures are presented at the maximum possible resolution.
Hence, we have uploaded the full dataset~\cite{akram_data}, allowing the reviewer to further analyze the cases in detail. 
Based on the results of Table \ref{tab:destine_hil}, the performance optimization of DESTinE can be attributed to several key factors: the use of graph-based structures that reduce network path computation complexity, parallelized event-driven simulation that enables efficient packet processing, dynamic topology adjustments that optimize network flow based on risk evaluation, and asynchronous server integration that enhances data transmission performance.

DESTinE operates on a discrete-event, event-driven architecture, where the system state is updated only when events occur, effectively skipping inactive periods. This significantly reduces unnecessary computations and enhances overall simulation speed. Furthermore, when DESTinE is operated in standalone mode, only virtual packets are generated, which are collected and analyzed exclusively by DESTinE’s \textbf{Sink} and \textbf{Port Monitor} classes. This internal handling of packet flow eliminates external processing overhead and significantly contributes to DESTinE’s execution speed, as reflected in Table \ref{tab:destine_hil} These combined optimizations enhance simulation efficiency and scalability, enabling DESTinE to handle large-scale power system networks with minimal computational overhead.

\section{Conclusion and Future Work}
\label{conclusion}

A discrete event simulation tool 
\name is developed and demonstrated in this paper to improve the scalability and efficiency in modeling large-scale cyber-physical systems.
\name is employed to 
analyze large-scale power system test cases ranging from 500 to 10,000-bus systems. The results highlight critical weak points in the network, especially under DoS attacks, and offer 
insights for improving network resilience through classification and ranking of nodes. 
\name allows 
us to accurately rank the critical 
routers and ensures reliable network performance even under
disturbances.
It also provides a resilient optimization capability, with particular emphasis on study of adverse conditions, and it allows us to accurately rank the critical routers and ensure reliable network performance even under disturbances. \name's capability to simulate both normal operations and adversarial conditions positions it as a 
platform for 
testing and optimizing large-scale cyber-physical infrastructures against cyber threats.


Future work will expand DESTinE's capabilities by incorporating adversarial scenarios such as data integrity breaches, false data injection, man-in-the-middle attacks, side-channel exploits, and ransomware attacks to broaden its cybersecurity applications. Enhancements will also focus on supporting flexible input formats and adaptive simulation parameters to accommodate diverse power system network topologies. Additionally, we plan to integrate DESTinE with existing cybersecurity frameworks to improve grid resilience and develop advanced anomaly detection algorithms to strengthen its ability to identify and mitigate cyber threats. While this study models the physical aspect through a network of routers derived from the energy infrastructure topology, future efforts will build on this foundation by incorporating a more comprehensive energy management framework, further enhancing the realism and applicability of DESTinE.

It should also be noted that DESTinE is primarily designed to focus on large-scale
cyber-physical interactions. 
Electromagnetic transient (EMT) analysis is not currently supported.
It is recognized that EMT phenomena can cause load fluctuations, which may lead to misclassification or unnecessary alerts that may impact 
certain anomaly detection scenarios, as described in \cite{Jena2024}. Therefore, the integration of EMT considerations into DESTinE is planned for future work to enable a more comprehensive analysis of the physical system, ensuring improved accuracy and robustness in cybersecurity assessments and anomaly detection.

\section*{Appendix}
\label{appendix}


The pseudocode for the generation of star and radial networks, along with the corresponding results presented in tabular form, is provided in this section.

\RestyleAlgo{ruled}

\renewcommand{\algorithmcfname}{Pseudo-code}

\begin{algorithm}
\scriptsize
\caption{Initialize Packet Generators and Routers for Substations (Star Network)} \label{algo1}
\SetAlgoLined
\KwData{List of substations, number of utilities}
\KwResult{Packet generators and routers setup for utilities}
\For{$i \gets 0$ \textbf{to} \texttt{len(number\_of\_utilities)}}{
    \For{\texttt{j in substations}}{
        \If{\texttt{"Region.Utility i"} \textbf{in} \texttt{j}}{
            Create a packet generator for substation \texttt{j} under \texttt{Utility i}: \\
            \texttt{pg\_substations["Utility i"][j] = PacketGenerator(SimPy.env, j)} \\
            
            Create a router for substation \texttt{j} under \texttt{Utility i}: \\
            \texttt{router\_substation["Utility i"][j] = Router(SimPy.env)} \\
        }
    }
}
\end{algorithm}

\begin{algorithm}
\scriptsize
\caption{Connect Packet Generators of Substations to Routers of Substations (Star Network)}
\label{algo2}
\SetAlgoLined
\KwData{List of substations, number of utilities}
\KwResult{Connections between packet generators and routers}

\For{$i \gets 0$ \textbf{to} \texttt{len(number\_of\_utilities)}}{
    \For{\texttt{j in substations}}{
        \If{\texttt{"Region.Utility i"} \textbf{in} \texttt{j}}{
            Connect the packet generator to the router: \\
            \texttt{pg\_substations["Utility i"][j].out = router\_substation["Utility i"][j]} \\
        }
    }
}
\end{algorithm}

\begin{algorithm}
\scriptsize
\caption{Connect Substation Routers to Utility Routers (Star Network)}
\label{algo3}
\SetAlgoLined
\KwData{List of substations, number of utilities}
\KwResult{Connections from substation router to utility routers}

\For{$i \gets 0$ \textbf{to} \texttt{len(number\_of\_utilities)}}{
    \For{\texttt{j in substations}}{
        \If{\texttt{"Region.Utility i"} \textbf{in} \texttt{j}}{
            Connect the substation's router to the utility's router: \\
            \texttt{router\_substation["Utility i"][j].out = router\_utility["Utility i"]} \\
        }
    }
}
\end{algorithm}


\begin{algorithm}
\scriptsize
\caption{Initialize Packet Generators and Routers for Initial Generation Substations (Radial Network)}
\label{algo4}
\SetAlgoLined
\KwData{Gen\_subs, Trans\_subs\_btw (Transmission Substation in between Generation Substation and Utility)}
\KwResult{Packet generators and routers setup}

\For{\texttt{sub} \textbf{in} \texttt{Gen\_subs}}{
    Create packet generator for \texttt{sub}: \\
    \texttt{pg\_gen[sub] = PacketGenerator(SimPy.env, sub)} \\
    Create routers for \texttt{sub}: \\
    \texttt{router\_gen[sub] = Router(SimPy.env)} \\
    
    Initialize transmission substation connection: \\
    \texttt{trans\_btw =  Trans\_subs\_btw[Gen\_subs.index(sub)]} \\
    Connect packet generator to transmission substation in between: \\
    \texttt{router\_gen[sub].out =} \\
    \texttt{router\_trans\_btw[trans\_btw]}
}
\end{algorithm}

\begin{algorithm}
\scriptsize
\caption{Identify and Connect Unconnected Routers to Utilities (Radial Network)}
\label{algo5}
\SetAlgoLined
\KwData{Gen\_subs, router\_trans\_btw}
\KwResult{Connected routers to utilities}

Identify \texttt{unconnected} as: \\
\texttt{[sub \textbf{for} sub \textbf{in} Gen\_subs} \\
\texttt{\textbf{if} router\_gen[sub].out \textbf{not} \textbf{in} vals\_trans\_btw]}

\For{\texttt{i} \textbf{in} \texttt{range(len(number\_of\_utilities))}}{
    \texttt{utility\_str = "Utility {i}"} \\
    \For{\texttt{sub} \textbf{in} \texttt{unconnected}}{
        \If{\texttt{utility\_str \textbf{in} sub}}{
            Connect unconnected router: \\
			            \texttt{temp1 = unconnected.index(sub)}\\
			          Comment:  Substation not in Generation Substations and Transmission substations in between Utilities\\
			            \texttt{temp2 =}\\ \texttt{mul\_sub\_trans\_btw[temp1]}\\            
            \texttt{router\_gen[sub].out =} \\

            \texttt{router\_trans\_btw[temp2]}
        }
    }
}
\end{algorithm}

\begin{algorithm}
\scriptsize
\caption{Setup Packet Generators and Routers for Transmission Substations (Radial Network)}
\label{algo7}
\SetAlgoLined
\KwData{Trans\_subs, Utilities}
\KwResult{Packet generators and routers for transmission substations}

\For{\texttt{i} \textbf{in} \texttt{range(len(number\_of\_utilities))}}{
    \texttt{utility\_str = "Utility {i}"} \\
    \For{\texttt{sub} \textbf{in} \texttt{Trans\_subs[utility\_str]}}{
        Create packet generator and switch port for \texttt{sub}: \\
        \texttt{pg\_trans[utility\_str][sub] =} \\
        \texttt{PacketGenerator(SimPy.env, sub, adist1, sdist)} \\
        \texttt{router\_trans[utility\_str][sub] =} \\
        \texttt{Router(SimPy.env, port\_rate)} \\
        
        Connect transmission router to utility router: \\
        \texttt{router\_trans[utility\_str][sub].out =} \\
        \texttt{router\_utility[utility\_str]} \\
        \texttt{pg\_trans[utility\_str][sub].out =} \\
        \texttt{sp\_trans[utility\_str][sub]}
    }
}
\end{algorithm}

\begin{table}[!hbt]
\centering
\captionsetup{justification=justified}
\caption{Ranking of critical utility routers for ACTIVSg500-bus case: Simulation vs Centrality Metrics, where Sim indicates the rank with discrete event simulation tool, $\mathbb{C}_B$, $\mathbb{X}$ and $\mathbb{C}_C$ denote the rank based on betweenness centrality, eigenvector centrality and closeness centrality matrix, respectively}
\label{tab:delay_normal_500}
\resizebox{\linewidth}{!}{%
\begin{tabular}{|>{\hspace{0pt}}m{0.173\linewidth}|>{\hspace{0pt}}m{0.108\linewidth}|>{\hspace{0pt}}m{0.088\linewidth}|>{\hspace{0pt}}m{0.052\linewidth}|>{\hspace{0pt}}m{0.088\linewidth}|>{\hspace{0pt}}m{0.119\linewidth}|>{\hspace{0pt}}m{0.1\linewidth}|>{\hspace{0pt}}m{0.06\linewidth}|>{\hspace{0pt}}m{0.1\linewidth}|} 
\hline
\textbf{Utility} \par{}\textbf{Router } & \textbf{Sim } & \textbf{ $\mathbb{C}_B$ } & \textbf{ $\mathbb{X}$  } & \textbf{ $\mathbb{C}_C$ } & \textbf{Sim } & \textbf{ $\mathbb{C}_B$ } & \textbf{ $\mathbb{X}$ } & \textbf{ $\mathbb{C}_C$} \\ 
\hline
\multicolumn{5}{|>{\centering\hspace{0pt}}m{0.509\linewidth}|}{\textbf{Star Topology }} & \multicolumn{4}{>{\centering\arraybackslash\hspace{0pt}}m{0.379\linewidth}|}{\textbf{Radial Topology}} \\ 
\hline
0 & 1 & 0 & 0 & 0 & 0 & 0 & 0 & 0 \\ 
\hline
1 & 0 & 2 & 2 & 2 & 1 & 2 & 2 & 2 \\ 
\hline
2 & 3 & 3 & 3 & 3 & 3 & 3 & 3 & 3 \\ 
\hline
3 & 2 & 1 & 1 & 1 & 2 & 1 & 1 & 1 \\
\hline
\end{tabular}
}
\end{table}

\begin{table}[!hbt]
\centering
\captionsetup{justification=justified}
\caption{Ranking of critical utility routers for ACTIVSg2000-bus case: Simulation vs Centrality Metrics, where  Sim indicates the rank with discrete event simulation tool, $\mathbb{C}_B$, $\mathbb{X}$, and $\mathbb{C}_C$ denote the rank based on betweenness centrality, eigenvector centrality and closeness centrality matrix, respectively}
\label{tab:delay_normal_2k}
\resizebox{\linewidth}{!}{%
\begin{tabular}{|>{\hspace{0pt}}m{0.169\linewidth}|>{\hspace{0pt}}m{0.106\linewidth}|>{\hspace{0pt}}m{0.087\linewidth}|>{\hspace{0pt}}m{0.071\linewidth}|>{\hspace{0pt}}m{0.087\linewidth}|>{\hspace{0pt}}m{0.11\linewidth}|>{\hspace{0pt}}m{0.092\linewidth}|>{\hspace{0pt}}m{0.075\linewidth}|>{\hspace{0pt}}m{0.092\linewidth}|} 
\hline
\textbf{Utility} \par{}\textbf{Router } & \textbf{Sim } & \textbf{$\mathbb{C}_B$} & \textbf{$\mathbb{X}$} & \textbf{$\mathbb{C}_C$} & \textbf{Sim } & \textbf{$\mathbb{C}_B$} & \textbf{$\mathbb{X}$} & \textbf{$\mathbb{C}_C$} \\ 
\hline
\multicolumn{5}{|>{\centering\hspace{0pt}}m{0.52\linewidth}|}{\textbf{Star Topology }} & \multicolumn{4}{>{\centering\arraybackslash\hspace{0pt}}m{0.369\linewidth}|}{\textbf{Radial Topology}} \\ 
\hline
2 & 4 & 4 & 4 & 2 & 2 & 4 & 4 & 4 \\ 
\hline
3 & 2 & 2 & 2 & 3 & 3 & 2 & 2 & 2 \\ 
\hline
0 & 0 & 0 & 0 & 0 & 0 & 0 & 0 & 0 \\ 
\hline
5 & 7 & 7 & 7 & 5 & 5 & 7 & 7 & 7 \\ 
\hline
11 & 15 & 15 & 15 & 11 & 11 & 15 & 15 & 15 \\ 
\hline
7 & 6 & 6 & 6 & 7 & 7 & 6 & 6 & 6 \\ 
\hline
8 & 10 & 10 & 10 & 8 & 8 & 10 & 9 & 10 \\ 
\hline
18 & 18 & 18 & 18 & 18 & 18 & 18 & 18 & 18 \\ 
\hline
14 & 11 & 11 & 11 & 14 & 14 & 11 & 10 & 11 \\ 
\hline
1 & 1 & 1 & 1 & 1 & 1 & 1 & 1 & 1 \\ 
\hline
4 & 3 & 3 & 3 & 4 & 4 & 3 & 3 & 3 \\ 
\hline
15 & 14 & 14 & 14 & 15 & 15 & 14 & 14 & 14 \\ 
\hline
10 & 8 & 8 & 8 & 10 & 10 & 8 & 8 & 8 \\ 
\hline
16 & 16 & 16 & 16 & 16 & 16 & 16 & 16 & 16 \\ 
\hline
13 & 9 & 9 & 9 & 13 & 13 & 9 & 12 & 9 \\ 
\hline
17 & 17 & 17 & 17 & 17 & 17 & 17 & 17 & 17 \\ 
\hline
12 & 13 & 13 & 13 & 12 & 12 & 13 & 11 & 13 \\ 
\hline
9 & 12 & 12 & 12 & 9 & 9 & 12 & 13 & 12 \\ 
\hline
6 & 5 & 5 & 5 & 6 & 6 & 5 & 5 & 5 \\ 
\hline
19 & 19 & 19 & 19 & 19 & 19 & 19 & 19 & 19 \\
\hline
\end{tabular}
}
\end{table}

\begin{table}[!hbt]
\centering
\captionsetup{justification=justified}
\caption{Ranking of critical utility routers for ACTIVSg10k-bus case: Simulation vs Centrality Metrics, where Sim indicates the rank with discrete event simulation tool, $\mathbb{C}_B$, $\mathbb{X}$, and $\mathbb{C}_C$ denote the rank based on betweenness centrality, eigenvector centrality and closeness centrality matrix, respectively}
\label{tab:delay_normal_west_us}
\resizebox{\linewidth}{!}{%
\begin{tabular}{|>{\hspace{0pt}}m{0.169\linewidth}|>{\hspace{0pt}}m{0.106\linewidth}|>{\hspace{0pt}}m{0.087\linewidth}|>{\hspace{0pt}}m{0.071\linewidth}|>{\hspace{0pt}}m{0.087\linewidth}|>{\hspace{0pt}}m{0.11\linewidth}|>{\hspace{0pt}}m{0.092\linewidth}|>{\hspace{0pt}}m{0.075\linewidth}|>{\hspace{0pt}}m{0.092\linewidth}|} 
\hline
\textbf{ Utility }\par{}\textbf{Router } & \textbf{ Sim } & \textbf{$\mathbb{C}_B$} & \textbf{$\mathbb{X}$ } & \textbf{$\mathbb{C}_C$ } & \textbf{ Sim } & \textbf{ $\mathbb{C}_B$ } & \textbf{$\mathbb{X}$} & \textbf{$\mathbb{C}_C$} \\ 
\hline
\multicolumn{5}{|>{\centering\hspace{0pt}}m{0.52\linewidth}|}{\textbf{ Star Topology }} & \multicolumn{4}{>{\centering\arraybackslash\hspace{0pt}}m{0.369\linewidth}|}{\textbf{ Radial Topology }} \\ 
\hline
0 & 5 & 3 & 6 & 6 & 5 & 3 & 6 & 6 \\ 
\hline
1 & 13 & 19 & 13 & 17 & 31 & 19 & 13 & 17 \\ 
\hline
2 & 25 & 69 & 65 & 63 & 73 & 69 & 65 & 63 \\ 
\hline
3 & 12 & 38 & 33 & 43 & 37 & 38 & 38 & 43 \\ 
\hline
4 & 24 & 22 & 44 & 29 & 6 & 22 & 41 & 32 \\ 
\hline
5 & 16 & 26 & 26 & 34 & 21 & 26 & 28 & 33 \\ 
\hline
6 & 9 & 14 & 31 & 15 & 14 & 14 & 30 & 15 \\ 
\hline
7 & 19 & 0 & 0 & 0 & 7 & 0 & 0 & 0 \\ 
\hline
8 & 64 & 43 & 29 & 33 & 68 & 43 & 27 & 31 \\ 
\hline
9 & 56 & 62 & 69 & 66 & 43 & 62 & 69 & 65 \\ 
\hline
10 & 11 & 53 & 49 & 51 & 15 & 53 & 49 & 55 \\ 
\hline
11 & 67 & 57 & 55 & 53 & 57 & 57 & 54 & 51 \\ 
\hline
12 & 18 & 12 & 9 & 9 & 29 & 12 & 9 & 9 \\ 
\hline
13 & 53 & 54 & 42 & 48 & 58 & 54 & 43 & 50 \\ 
\hline
14 & 23 & 23 & 14 & 18 & 20 & 23 & 14 & 19 \\ 
\hline
15 & 10 & 9 & 21 & 13 & 10 & 10 & 23 & 13 \\ 
\hline
16 & 20 & 24 & 37 & 30 & 12 & 24 & 32 & 30 \\ 
\hline
17 & 68 & 72 & 74 & 73 & 71 & 72 & 74 & 73 \\ 
\hline
18 & 31 & 52 & 54 & 56 & 35 & 52 & 57 & 56 \\ 
\hline
19 & 27 & 56 & 60 & 60 & 23 & 56 & 61 & 60 \\ 
\hline
20 & 6 & 25 & 24 & 32 & 2 & 25 & 37 & 34 \\ 
\hline
21 & 60 & 65 & 71 & 70 & 65 & 65 & 71 & 69 \\ 
\hline
22 & 32 & 49 & 62 & 58 & 33 & 48 & 62 & 58 \\ 
\hline
23 & 49 & 46 & 52 & 49 & 46 & 46 & 52 & 47 \\ 
\hline
24 & 3 & 4 & 7 & 7 & 22 & 4 & 7 & 7 \\ 
\hline
25 & 0 & 16 & 19 & 22 & 3 & 16 & 19 & 21 \\ 
\hline
26 & 43 & 35 & 46 & 40 & 32 & 35 & 46 & 40 \\ 
\hline
27 & 69 & 66 & 59 & 64 & 63 & 66 & 59 & 64 \\ 
\hline
28 & 26 & 29 & 30 & 35 & 28 & 29 & 29 & 35 \\ 
\hline
29 & 70 & 58 & 56 & 54 & 59 & 58 & 56 & 52 \\ 
\hline
30 & 22 & 20 & 20 & 27 & 13 & 20 & 24 & 27 \\ 
\hline
31 & 30 & 31 & 50 & 45 & 11 & 31 & 50 & 45 \\ 
\hline
32 & 55 & 70 & 73 & 72 & 62 & 70 & 73 & 72 \\ 
\hline
33 & 46 & 47 & 41 & 47 & 38 & 47 & 42 & 48 \\ 
\hline
34 & 78 & 73 & 66 & 67 & 78 & 73 & 66 & 66 \\ 
\hline
35 & 44 & 39 & 39 & 36 & 30 & 39 & 35 & 36 \\ 
\hline
36 & 2 & 2 & 2 & 2 & 0 & 2 & 2 & 2 \\ 
\hline
37 & 50 & 27 & 36 & 19 & 41 & 27 & 33 & 18 \\ 
\hline
38 & 4 & 10 & 22 & 14 & 4 & 9 & 21 & 14 \\ 
\hline
39 & 39 & 68 & 67 & 65 & 45 & 68 & 67 & 70 \\ 
\hline
40 & 79 & 78 & 78 & 79 & 79 & 78 & 78 & 78 \\ 
\hline
41 & 37 & 28 & 15 & 21 & 39 & 28 & 15 & 22 \\ 
\hline
42 & 33 & 6 & 3 & 3 & 19 & 6 & 5 & 3 \\ 
\hline
43 & 71 & 77 & 77 & 77 & 69 & 77 & 77 & 77 \\ 
\hline
44 & 45 & 33 & 16 & 24 & 52 & 33 & 16 & 23 \\ 
\hline
45 & 36 & 60 & 64 & 61 & 50 & 60 & 64 & 61 \\ 
\hline
46 & 66 & 67 & 51 & 68 & 64 & 67 & 51 & 67 \\ 
\hline
47 & 14 & 7 & 8 & 8 & 27 & 7 & 8 & 8 \\ 
\hline
48 & 8 & 15 & 23 & 23 & 9 & 15 & 20 & 26 \\ 
\hline
49 & 21 & 48 & 61 & 57 & 25 & 49 & 60 & 57 \\ 
\hline
50 & 1 & 5 & 12 & 12 & 1 & 5 & 12 & 12 \\ 
\hline
51 & 61 & 50 & 35 & 46 & 66 & 50 & 40 & 46 \\ 
\hline
52 & 7 & 1 & 1 & 1 & 8 & 1 & 1 & 1 \\ 
\hline
53 & 48 & 41 & 47 & 41 & 36 & 41 & 47 & 41 \\ 
\hline
54 & 77 & 76 & 76 & 75 & 75 & 76 & 75 & 75 \\ 
\hline
55 & 42 & 32 & 45 & 39 & 24 & 32 & 45 & 39 \\ 
\hline
56 & 59 & 64 & 70 & 69 & 51 & 64 & 70 & 68 \\ 
\hline
57 & 41 & 55 & 63 & 59 & 48 & 55 & 63 & 59 \\ 
\hline
58 & 73 & 61 & 43 & 52 & 70 & 61 & 44 & 54 \\ 
\hline
59 & 58 & 45 & 32 & 38 & 54 & 45 & 31 & 38 \\ 
\hline
60 & 28 & 18 & 10 & 10 & 26 & 18 & 10 & 10 \\ 
\hline
61 & 54 & 36 & 17 & 26 & 60 & 36 & 17 & 25 \\ 
\hline
62 & 57 & 34 & 38 & 20 & 55 & 34 & 34 & 20 \\ 
\hline
63 & 72 & 75 & 68 & 76 & 72 & 75 & 68 & 76 \\ 
\hline
64 & 17 & 17 & 25 & 25 & 18 & 17 & 22 & 24 \\ 
\hline
65 & 62 & 44 & 48 & 42 & 56 & 44 & 48 & 42 \\ 
\hline
66 & 75 & 74 & 75 & 74 & 74 & 74 & 76 & 74 \\ 
\hline
67 & 65 & 59 & 57 & 55 & 67 & 59 & 55 & 53 \\ 
\hline
68 & 51 & 40 & 40 & 37 & 53 & 40 & 36 & 37 \\ 
\hline
69 & 34 & 42 & 34 & 44 & 42 & 42 & 39 & 44 \\ 
\hline
70 & 76 & 79 & 79 & 78 & 77 & 79 & 79 & 79 \\ 
\hline
71 & 15 & 13 & 18 & 16 & 17 & 13 & 18 & 16 \\ 
\hline
72 & 29 & 8 & 4 & 4 & 34 & 8 & 4 & 4 \\ 
\hline
73 & 63 & 63 & 58 & 62 & 61 & 63 & 58 & 62 \\ 
\hline
74 & 35 & 11 & 5 & 5 & 16 & 11 & 3 & 5 \\ 
\hline
75 & 47 & 51 & 53 & 50 & 49 & 51 & 53 & 49 \\ 
\hline
76 & 38 & 21 & 11 & 11 & 40 & 21 & 11 & 11 \\ 
\hline
77 & 52 & 37 & 28 & 31 & 47 & 37 & 26 & 29 \\ 
\hline
78 & 74 & 71 & 72 & 71 & 76 & 71 & 72 & 71 \\ 
\hline
79 & 40 & 30 & 27 & 28 & 44 & 30 & 25 & 28 \\
\hline
\end{tabular}
}
\end{table}

\begin{table}[!hbt]
\centering
\captionsetup{justification=justified}
\caption{Impact of DoS threat for ACTIVSg500 case with star and radial topology (NA denotes Not Applicable; for the "Substations Lost (Radial)" column, only the substations directly connected to the utility are considered)}
\label{tab:dos_500}
\resizebox{\linewidth}{!}{%
\begin{tabular}{|>{\hspace{0pt}}m{0.144\linewidth}|>{\hspace{0pt}}m{0.094\linewidth}|>{\hspace{0pt}}m{0.1\linewidth}|>{\hspace{0pt}}m{0.102\linewidth}|>{\hspace{0pt}}m{0.102\linewidth}|>{\hspace{0pt}}m{0.194\linewidth}|>{\hspace{0pt}}m{0.194\linewidth}|} 
\hline
\textbf{ Utility No. }\par{}\textbf{Affected by  }\par{}\textbf{DoS } & \textbf{ $\alpha_0$ }\par{}\textbf{(Star) } & \textbf{ $\alpha_1$ }\par{}\textbf{(Star) } & \textbf{ $\alpha_0$ }\par{}\textbf{(Radial) } & \textbf{ $\alpha_1$ }\par{}\textbf{(Radial) } & \textbf{ Substation Lost }\par{}\textbf{(Star) } & \textbf{ Substation Lost }\par{}\textbf{(Radial) } \\ 
\hline
0 & \multirow{5}{0.094\linewidth}{\hspace{0pt}0.1} & \multirow{5}{0.1\linewidth}{\hspace{0pt}0.1} & \multirow{5}{0.102\linewidth}{\hspace{0pt}0.1} & \multirow{5}{0.102\linewidth}{\hspace{0pt}0.0} & 67 & 57 \\ 
\cline{1-1}\cline{6-7}
1 &  &  &  &  & 50 & 42 \\ 
\cline{1-1}\cline{6-7}
2 &  &  &  &  & 39 & 36 \\ 
\cline{1-1}\cline{6-7}
3 &  &  &  &  & 52 & 43 \\ 
\cline{1-1}\cline{6-7}
Stable \par{}(No DoS) &  &  &  &  & NA & NA \\
\hline
\end{tabular}
}
\end{table}

\begin{table}[!hbt]
\centering
\captionsetup{justification=justified}
\caption{Classification of severity on the impact of DoS threat for ACTIVSg2000 case with star topology (NA denotes Not Applicable)}
\label{tab:dos_2k_star}
\resizebox{\linewidth}{!}{%
\begin{tabular}{|>{\hspace{0pt}}m{0.44\linewidth}|>{\hspace{0pt}}m{0.158\linewidth}|>{\hspace{0pt}}m{0.096\linewidth}|>{\hspace{0pt}}m{0.096\linewidth}|>{\hspace{0pt}}m{0.165\linewidth}|} 
\hline
\textbf{Status} & \textbf{Utility No.} \par{}\textbf{Affected by DoS} & \textbf{$\alpha_0$} & \textbf{$\alpha_1$} & \textbf{Substations} \par{}\textbf{Lost} \\ 
\hline
\multirow{4}{0.44\linewidth}{\hspace{0pt}\textbf{Catastrophic (6)}} & 8 & 0.85 & 0.30 & 45 \\ 
\cline{2-5}
 & 9 & 0.60 & 0.20 & 163 \\ 
\cline{2-5}
 & 10 & 0.75 & 0.25 & 103 \\ 
\cline{2-5}
 & 16 & 0.95 & 0.30 & 40 \\ 
\hline
\multirow{3}{0.44\linewidth}{\hspace{0pt}\textbf{Severe (5)}} & 3 & 0.75 & 0.25 & 57 \\ 
\cline{2-5}
 & 4 & 0.70 & 0.25 & 34 \\ 
\cline{2-5}
 & 6 & 0.70 & 0.25 & 45 \\ 
\hline
\multirow{3}{0.44\linewidth}{\hspace{0pt}\textbf{Critical (4)}} & 0 & 0.30 & 0.25 & 98 \\ 
\cline{2-5}
 & 2 & 0.30 & 0.10 & 199 \\ 
\cline{2-5}
 & 18 & 0.65 & 0.20 & 65 \\ 
\hline
\multirow{3}{0.44\linewidth}{\hspace{0pt}\textbf{High (3)}} & 12 & 0.45 & 0.15 & 48 \\ 
\cline{2-5}
 & 14 & 0.40 & 0.15 & 47 \\ 
\cline{2-5}
 & 15 & 0.45 & 0.15 & 20 \\ 
\hline
\multirow{3}{0.44\linewidth}{\hspace{0pt}\textbf{Extreme (2)}} & 1 & 0.30 & 0.10 & 104 \\ 
\cline{2-5}
 & 7 & 0.45 & 0.15 & 13 \\ 
\cline{2-5}
 & 17 & 0.35 & 0.10 & 41 \\ 
\hline
\multirow{4}{0.44\linewidth}{\hspace{0pt}\textbf{Elevated (1)}} & 5 & 0.25 & 0.10 & 60 \\ 
\cline{2-5}
 & 11 & 0.30 & 0.10 & 40 \\ 
\cline{2-5}
 & 13 & 0.25 & 0.10 & 22 \\ 
\cline{2-5}
 & 19 & 0.25 & 0.10 & 6 \\ 
\hline
\textbf{Normal} & NA & 0.25 & 0.10 & NA \\
\hline
\end{tabular}
}
\end{table}

\begin{table}[!hbt]
\centering
\captionsetup{justification=justified}
\caption{Classification of severity on the impact of DoS threat for ACTIVSg2000 case with radial topology (NA denotes Not Applicable; for the "Substations Lost" column, only the substations directly connected to the utility are considered)}
\label{tab:dos_2k_radial}
\resizebox{\linewidth}{!}{%
\begin{tabular}{|>{\hspace{0pt}}m{0.44\linewidth}|>{\hspace{0pt}}m{0.158\linewidth}|>{\hspace{0pt}}m{0.096\linewidth}|>{\hspace{0pt}}m{0.096\linewidth}|>{\hspace{0pt}}m{0.165\linewidth}|} 
\hline
\textbf{Status} & \textbf{Utility No.} \par{}\textbf{Affected by DoS} & \textbf{$\alpha_0$} & \textbf{$\alpha_1$} & \textbf{Substations} \par{}\textbf{Lost} \\ 
\hline
\multirow{4}{0.44\linewidth}{\hspace{0pt}\textbf{Catastrophic (6)}} & 1 & 0.90 & 0.30 & 102 \\ 
\cline{2-5}
 & 3 & 0.85 & 0.30 & 37 \\ 
\cline{2-5}
 & 8 & 0.85 & 0.30 & 40 \\ 
\cline{2-5}
 & 12 & 0.60 & 0.20 & 188 \\ 
\hline
\multirow{3}{0.44\linewidth}{\hspace{0pt}\textbf{Severe (5)}} & 2 & 0.65 & 0.20 & 99 \\ 
\cline{2-5}
 & 4 & 0.65 & 0.20 & 40 \\ 
\cline{2-5}
 & 11 & 0.70 & 0.25 & 6 \\ 
\hline
\multirow{3}{0.44\linewidth}{\hspace{0pt}\textbf{Critical (4)}} & 5 & 0.60 & 0.20 & 20 \\ 
\cline{2-5}
 & 6 & 0.55 & 0.20  & 39 \\ 
\cline{2-5}
 & 15 & 0.55 & 0.20 & 56 \\ 
\hline
\multirow{3}{0.44\linewidth}{\hspace{0pt}\textbf{High (3)}} & 9 & 0.55 & 0.15 & 39 \\ 
\cline{2-5}
 & 18 & 0.55 & 0.15 & 39 \\ 
\cline{2-5}
 & 19 & 0.30 & 0.10 & 158 \\ 
\hline
\multirow{3}{0.44\linewidth}{\hspace{0pt}\textbf{Extreme (2)}} & 7 & 0.50 & 0.15 & 19 \\ 
\cline{2-5}
 & 10 & 0.40 & 0.15 & 64 \\ 
\cline{2-5}
 & 13 & 0.45 & 0.15 & 51 \\ 
\hline
\multirow{4}{0.44\linewidth}{\hspace{0pt}\textbf{Elevated (1)}} & 0 & 0.35  & 0.10  & 97 \\ 
\cline{2-5}
 & 14 & 0.25 & 0.10 & 29 \\ 
\cline{2-5}
 & 16 & 0.30 & 0.10 & 43 \\ 
\cline{2-5}
 & 17 & 0.30  & 0.10 & 13 \\ 
\hline
\textbf{Normal} & NA & 0.30 & 0.10 & NA \\
\hline
\end{tabular}
}
\end{table}

\begin{table}[!hbt]
\centering
\captionsetup{justification=justified}
\caption{Classification of severity on the impact of DoS threat for ACTIVSg10k case with star topology (NA denotes Not Applicable)}
\label{tab:dos_star}
\resizebox{\linewidth}{!}{%
\begin{tabular}{|>{\hspace{0pt}}m{0.44\linewidth}|>{\hspace{0pt}}m{0.145\linewidth}|>{\hspace{0pt}}m{0.046\linewidth}|>{\hspace{0pt}}m{0.046\linewidth}|>{\hspace{0pt}}m{0.165\linewidth}|>{\hspace{0pt}}m{0.16\linewidth}|} 
\hline
\textbf{Status} & \textbf{Utility No.} \par{}\textbf{Affected by~DoS} & \textbf{$\alpha_0$} & \textbf{$\alpha_1$} & \textbf{Substations} \par{}\textbf{Lost} & \textbf{Regulatory No.} \par{}\textbf{Affected} \\ 
\hline
\multirow{14}{0.44\linewidth}{\hspace{0pt}\textbf{Catastrophic (6)}} & 0 & 4.0 & 1.5 & 232 & 14 \\ 
\cline{2-6}
 & 4 & 4.0 & 1.5 & 107 & 10 \\ 
\cline{2-6}
 & 7 & 4.0 & 1.5 & 298 & 6 \\ 
\cline{2-6}
 & 15 & 4.0 & 1.5 & 135 & 13 \\ 
\cline{2-6}
 & 24 & 4.0 & 1.5 & 137 & 14 \\ 
\cline{2-6}
 & 36 & 4.0 & 1.5 & 151 & 6 \\ 
\cline{2-6}
 & 38 & 4.0 & 1.5 & 135 & 13 \\ 
\cline{2-6}
 & 44 & 4.5 & 1.5 & 29 & 3 \\ 
\cline{2-6}
 & 48 & 4.0 & 1.5 & 136 & 8 \\ 
\cline{2-6}
 & 50 & 4.5 & 1.5 & 181 & 3 \\ 
\cline{2-6}
 & 52 & 4.0 & 1.5 & 226 & 6 \\ 
\cline{2-6}
 & 55 & 4.5 & 1.5 & 48 & 10 \\ 
\cline{2-6}
 & 64 & 4.0 & 1.5 & 100 & 4 \\ 
\cline{2-6}
 & 71 & 5.0 & 2.0 & 143 & 4 \\ 
\hline
\multirow{13}{0.44\linewidth}{\hspace{0pt}\textbf{Severe (5)}} & 1 & 4.0 & 1.5 & 73 & 3 \\ 
\cline{2-6}
 & 3 & 4.0 & 1.5 & 43 & 19 \\ 
\cline{2-6}
 & 10 & 4.0 & 1.5 & 80 & 9 \\ 
\cline{2-6}
 & 12 & 4.0 & 1.5 & 67 & 14 \\ 
\cline{2-6}
 & 14 & 4.0 & 1.5 & 57 & 3 \\ 
\cline{2-6}
 & 16 & 4.0 & 1.5 & 84 & 8 \\ 
\cline{2-6}
 & 22 & 4.0 & 1.5 & 48 & 18 \\ 
\cline{2-6}
 & 26 & 4.0 & 1.5 & 42 & 10 \\ 
\cline{2-6}
 & 28 & 4.0 & 1.5 & 49 & 12 \\ 
\cline{2-6}
 & 39 & 4.0 & 1.5 & 44 & 1 \\ 
\cline{2-6}
 & 47 & 4.0 & 1.5 & 90 & 14 \\ 
\cline{2-6}
 & 49 & 4.0 & 1.5 & 48 & 18 \\ 
\cline{2-6}
 & 72 & 4.0 & 1.5 & 60 & 6 \\ 
\hline
\multirow{13}{0.44\linewidth}{\hspace{0pt}\textbf{Critical (4)}} & 9 & 4.0 & 1.5 & 29 & 2 \\ 
\cline{2-6}
 & 20 & 3.5 & 1.5 & 130 & 0 \\ 
\cline{2-6}
 & 33 & 4.0 & 1.5 & 30 & 0 \\ 
\cline{2-6}
 & 35 & 4.0 & 1.5 & 35 & 8 \\ 
\cline{2-6}
 & 37 & 4.0 & 1.5 & 32 & 13 \\ 
\cline{2-6}
 & 41 & 4.0 & 1.5 & 33 & 3 \\ 
\cline{2-6}
 & 45 & 4.0 & 1.5 & 35 & 15 \\ 
\cline{2-6}
 & 68 & 4.0 & 1.5 & 33 & 8 \\ 
\cline{2-6}
 & 69 & 4.0 & 1.5 & 35 & 19 \\ 
\cline{2-6}
 & 75 & 4.0 & 1.5 & 28 & 5 \\ 
\cline{2-6}
 & 76 & 4.0 & 1.5 & 40 & 14 \\ 
\cline{2-6}
 & 77 & 4.0 & 1.5 & 30 & 4 \\ 
\cline{2-6}
 & 79 & 4.0 & 1.5 & 39 & 4 \\ 
\hline
\multirow{13}{0.44\linewidth}{\hspace{0pt}\textbf{High (3)}} & 2 & 4.0 & 1.5 & 15 & 15 \\ 
\cline{2-6}
 & 8 & 4.0 & 1.5 & 21 & 4 \\ 
\cline{2-6}
 & 13 & 4.0 & 1.5 & 25 & 0 \\ 
\cline{2-6}
 & 17 & 4.0 & 1.5 & 16 & 17 \\ 
\cline{2-6}
 & 21 & 4.0 & 1.5 & 22 & 2 \\ 
\cline{2-6}
 & 29 & 4.0 & 1.5 & 19 & 5 \\ 
\cline{2-6}
 & 46 & 4.0 & 1.5 & 19 & 9 \\ 
\cline{2-6}
 & 51 & 4.0 & 1.5 & 23 & 19 \\ 
\cline{2-6}
 & 56 & 4.0 & 1.5 & 25 & 2 \\ 
\cline{2-6}
 & 59 & 4.0 & 1.5 & 23 & 12 \\ 
\cline{2-6}
 & 62 & 4.0 & 1.5 & 26 & 13 \\ 
\cline{2-6}
 & 67 & 4.0 & 1.5 & 18 & 5 \\ 
\cline{2-6}
 & 73 & 4.0 & 1.5 & 21 & 7 \\ 
\hline
\multirow{13}{0.44\linewidth}{\hspace{0pt}\textbf{Extreme (2)}} & 5 & 3.5 & 1.5 & 66 & 12 \\ 
\cline{2-6}
 & 6 & 3.5 & 1.5 & 87 & 13 \\ 
\cline{2-6}
 & 19 & 3.5 & 1.5 & 50 & 15 \\ 
\cline{2-6}
 & 27 & 3.5 & 1.5 & 18 & 7 \\ 
\cline{2-6}
 & 31 & 3.5 & 1.5 & 74 & 5 \\ 
\cline{2-6}
 & 32 & 3.5 & 1.5 & 24 & 17 \\ 
\cline{2-6}
 & 34 & 4.0 & 1.5 & 8 & 15 \\ 
\cline{2-6}
 & 40 & 4.0 & 1.5 & 7 & 11 \\ 
\cline{2-6}
 & 42 & 3.5 & 1.0 & 63 & 6 \\ 
\cline{2-6}
 & 63 & 4.0 & 1.5 & 14 & 1 \\ 
\cline{2-6}
 & 66 & 3.5 & 1.5 & 13 & 17 \\ 
\cline{2-6}
 & 70 & 4.0 & 1.5 & 11 & 16 \\ 
\cline{2-6}
 & 78 & 4.0 & 1.5 & 13 & 2 \\ 
\hline
\multirow{14}{0.44\linewidth}{\hspace{0pt}\textbf{Elevated (1)}} & 11 & 2.5 & 1.5 & 19 & 5 \\ 
\cline{2-6}
 & 18 & 3.0 & 1.0 & 66 & 7 \\ 
\cline{2-6}
 & 23 & 3.5 & 1.0 & 32 & 5 \\ 
\cline{2-6}
 & 25 & 2.5 & 1.0 & 142 & 12 \\ 
\cline{2-6}
 & 30 & 3.0 & 1.0 & 141 & 19 \\ 
\cline{2-6}
 & 43 & 3.0 & 1.0 & 17 & 11 \\ 
\cline{2-6}
 & 53 & 3.0 & 1.0 & 35 & 10 \\ 
\cline{2-6}
 & 54 & 3.0 & 1.0 & 10 & 17 \\ 
\cline{2-6}
 & 57 & 2.5 & 1.0 & 39 & 18 \\ 
\cline{2-6}
 & 58 & 2.5 & 1.0 & 14 & 0 \\ 
\cline{2-6}
 & 60 & 3.0 & 1.0 & 46 & 14 \\ 
\cline{2-6}
 & 61 & 3.5 & 1.0 & 25 & 3 \\ 
\cline{2-6}
 & 65 & 2.5 & 1.0 & 26 & 10 \\ 
\cline{2-6}
 & 74 & 2.5 & 1.0 & 57 & 6 \\ 
\hline
\textbf{Normal} & NA & 2.5 & 1.0 & NA & NA \\
\hline
\end{tabular}
}
\end{table}

\begin{table}[!hbt]
\centering
\centering
\captionsetup{justification=justified}
\caption{Classification of severity on the impact of DoS threat for ACTIVSg10k case with radial topology (NA denotes Not Applicable; for the "Substations Lost" column, only the substations directly connected to the utility are considered)}
\label{tab:dos_radial}
\resizebox{\linewidth}{!}{%
\begin{tabular}{|>{\hspace{0pt}}m{0.44\linewidth}|>{\hspace{0pt}}m{0.145\linewidth}|>{\hspace{0pt}}m{0.046\linewidth}|>{\hspace{0pt}}m{0.046\linewidth}|>{\hspace{0pt}}m{0.165\linewidth}|>{\hspace{0pt}}m{0.16\linewidth}|} 
\hline
\textbf{Status} & \textbf{Utility No.} \par{}\textbf{Affected by DoS}  & \textbf{$\alpha_0$} & \textbf{$\alpha_1$} & \textbf{Substations} \par{}\textbf{Lost} & \textbf{Regulatory No.} \par{}\textbf{Affected} \\ 
\hline
\multirow{14}{0.44\linewidth}{\hspace{0pt}\textbf{Catastrophic (6)}} & 0 & 4.0 & 1.5 & 214 & 14 \\ 
\cline{2-6}
 & 9 & 4.5 & 1.5 & 28 & 2 \\ 
\cline{2-6}
 & 14 & 4.5 & 1.5 & 57 & 3 \\ 
\cline{2-6}
 & 16 & 4.5 & 1.5 & 79 & 8 \\ 
\cline{2-6}
 & 21 & 4.5 & 1.5 & 17 & 2 \\ 
\cline{2-6}
 & 25 & 4.0 & 1.5 & 136 & 12 \\ 
\cline{2-6}
 & 31 & 4.5 & 1.5 & 70 & 5 \\ 
\cline{2-6}
 & 33 & 4.5 & 1.5 & 30 & 0 \\ 
\cline{2-6}
 & 36 & 4.0 & 1.5 & 140 & 6 \\ 
\cline{2-6}
 & 41 & 4.5 & 1.5 & 32 & 3 \\ 
\cline{2-6}
 & 43 & 4.5 & 1.5 & 14 & 11 \\ 
\cline{2-6}
 & 48 & 4.0 & 1.5 & 131 & 8 \\ 
\cline{2-6}
 & 71 & 4.0 & 1.5 & 140 & 4 \\ 
\cline{2-6}
 & 73 & 4.5 & 1.5 & 18 & 7 \\ 
\hline
\multirow{12}{0.44\linewidth}{\hspace{0pt}\textbf{Severe (5)}} & 1 & 4.0 & 1.5 & 68 & 3 \\ 
\cline{2-6}
 & 4 & 4.0 & 1.5 & 103 & 10 \\ 
\cline{2-6}
 & 5 & 4.0 & 1.5 & 69 & 12 \\ 
\cline{2-6}
 & 15 & 4.0 & 1.5 & 126 & 13 \\ 
\cline{2-6}
 & 18 & 4.0 & 1.5 & 58 & 7 \\ 
\cline{2-6}
 & 24 & 4.0 & 1.5 & 124 & 14 \\ 
\cline{2-6}
 & 42 & 4.0 & 1.5 & 51 & 6 \\ 
\cline{2-6}
 & 47 & 4.0 & 1.5 & 74 & 14 \\ 
\cline{2-6}
 & 50 & 3.5 & 1.5 & 173 & 3 \\ 
\cline{2-6}
 & 52 & 3.5 & 1.5 & 220 & 6 \\ 
\cline{2-6}
 & 72 & 4.0 & 1.5 & 48 & 6 \\ 
\cline{2-6}
 & 74 & 4.0 & 1.5 & 55 & 6 \\ 
\hline
\multirow{14}{0.44\linewidth}{\hspace{0pt}\textbf{Critical (4)}} & 3 & 4.0 & 1.5 & 35 & 19 \\ 
\cline{2-6}
 & 19 & 4.0 & 1.5 & 48 & 15 \\ 
\cline{2-6}
 & 26 & 4.0 & 1.5 & 35 & 10 \\ 
\cline{2-6}
 & 28 & 4.0 & 1.5 & 46 & 12 \\ 
\cline{2-6}
 & 35 & 4.0 & 1.5 & 34 & 8 \\ 
\cline{2-6}
 & 37 & 4.0 & 1.5 & 30 & 13 \\ 
\cline{2-6}
 & 44 & 4.0 & 1.5 & 21 & 3 \\ 
\cline{2-6}
 & 45 & 4.0 & 1.5 & 26 & 15 \\ 
\cline{2-6}
 & 57 & 4.0 & 1.5 & 35 & 18 \\ 
\cline{2-6}
 & 60 & 4.0 & 1.5 & 42 & 14 \\ 
\cline{2-6}
 & 68 & 4.0 & 1.5 & 25 & 8 \\ 
\cline{2-6}
 & 75 & 4.0 & 1.5 & 25 & 5 \\ 
\cline{2-6}
 & 77 & 4.0 & 1.5 & 27 & 4 \\ 
\cline{2-6}
 & 79 & 4.0 & 1.5 & 29 & 4 \\ 
\hline
\multirow{13}{0.44\linewidth}{\hspace{0pt}\textbf{High (3)}} & 2 & 4.0 & 1.5 & 13 & 15 \\ 
\cline{2-6}
 & 8 & 4.0 & 1.5 & 17 & 4 \\ 
\cline{2-6}
 & 11 & 4.0 & 1.5 & 20 & 5 \\ 
\cline{2-6}
 & 13 & 4.0 & 1.5 & 20 & 0 \\ 
\cline{2-6}
 & 17 & 4.0 & 1.5 & 13 & 17 \\ 
\cline{2-6}
 & 27 & 4.0 & 1.5 & 18 & 7 \\ 
\cline{2-6}
 & 46 & 4.0 & 1.5 & 17 & 9 \\ 
\cline{2-6}
 & 51 & 4.0 & 1.5 & 19 & 19 \\ 
\cline{2-6}
 & 58 & 4.0 & 1.5 & 24 & 0 \\ 
\cline{2-6}
 & 61 & 4.0 & 1.5 & 20 & 3 \\ 
\cline{2-6}
 & 62 & 4.0 & 1.5 & 22 & 13 \\ 
\cline{2-6}
 & 65 & 4.0 & 1.5 & 22 & 10 \\ 
\cline{2-6}
 & 67 & 4.0 & 1.5 & 18 & 5 \\ 
\hline
\multirow{13}{0.44\linewidth}{\hspace{0pt}\textbf{Extreme (2)}} & 7 & 2.5 & 1 & 279 & 6 \\ 
\cline{2-6}
 & 20 & 3.5 & 1.5 & 115 & 0 \\ 
\cline{2-6}
 & 23 & 3.5 & 1.5 & 30 & 5 \\ 
\cline{2-6}
 & 29 & 3.5 & 1.5 & 20 & 5 \\ 
\cline{2-6}
 & 30 & 3.5 & 1.5 & 125 & 19 \\ 
\cline{2-6}
 & 34 & 4.0 & 1.5 & 8 & 15 \\ 
\cline{2-6}
 & 40 & 4.0 & 1.5 & 7 & 11 \\ 
\cline{2-6}
 & 63 & 4.0 & 1.5 & 12 & 1 \\ 
\cline{2-6}
 & 64 & 3.5 & 1.5 & 96 & 4 \\ 
\cline{2-6}
 & 66 & 3.5 & 1.5 & 13 & 17 \\ 
\cline{2-6}
 & 69 & 3.5 & 1.5 & 29 & 19 \\ 
\cline{2-6}
 & 70 & 3.5 & 1.5 & 10 & 16 \\ 
\cline{2-6}
 & 78 & 4.0 & 1.5 & 10 & 2 \\ 
\hline
\multirow{14}{0.44\linewidth}{\hspace{0pt}\textbf{Elevated (1)}} & 6 & 2.5 & 1 & 80 & 13 \\ 
\cline{2-6}
 & 10 & 3.0 & 1 & 73 & 9 \\ 
\cline{2-6}
 & 12 & 2.5 & 1 & 57 & 14 \\ 
\cline{2-6}
 & 22 & 2.5 & 1 & 43 & 18 \\ 
\cline{2-6}
 & 32 & 3.0 & 1 & 20 & 17 \\ 
\cline{2-6}
 & 38 & 2.5 & 1 & 132 & 13 \\ 
\cline{2-6}
 & 39 & 2.5 & 1 & 37 & 1 \\ 
\cline{2-6}
 & 49 & 2.5 & 1 & 42 & 18 \\ 
\cline{2-6}
 & 53 & 3.0 & 1 & 33 & 10 \\ 
\cline{2-6}
 & 54 & 3.0 & 1 & 10 & 17 \\ 
\cline{2-6}
 & 55 & 2.5 & 1 & 47 & 10 \\ 
\cline{2-6}
 & 56 & 3.5 & 1 & 24 & 2 \\ 
\cline{2-6}
 & 59 & 3.5 & 1 & 24 & 12 \\ 
\cline{2-6}
 & 76 & 2.5 & 1 & 34 & 14 \\ 
\hline
\textbf{Normal} & NA & 2.5 & 1 & NA & NA \\
\hline
\end{tabular}
}
\end{table}

\clearpage
\bibliographystyle{IEEEtran}
\bibliography{ref}



\end{document}